\documentclass{ws-jmmb}
\usepackage{graphicx}
\usepackage{float}
\usepackage{amssymb}
\usepackage{subfigure}
\usepackage[super,sort&compress]{natbib}

\begin{document}

\markboth{Omer San and Anne E. Staples}{Reduced-order modeling of physiological fluid dynamics}

%
\catchline{}{}{}{}{}
%

\title{AN IMPROVED MODEL FOR REDUCED-ORDER PHYSIOLOGICAL FLUID FLOWS}

\author{Omer San and Anne E. Staples}

\address{Department of Engineering Science and Mechanics \\ Virginia Tech, Blacksburg, VA 24061, USA\\
E-mail addresses: omersan@vt.edu, aestaples@vt.edu}

\maketitle


\begin{abstract}
An improved one-dimensional mathematical model based on Pulsed Flow Equations (PFE) is derived by integrating the axial component of the momentum equation over the transient Womersley velocity profile, providing a dynamic momentum equation whose coefficients are smoothly varying functions of the spatial variable. The resulting momentum equation along with the continuity equation and pressure-area relation form our reduced-order model for physiological fluid flows in one dimension, and are aimed at providing accurate and fast-to-compute global models for physiological systems represented as networks of quasi one-dimensional fluid flows. The consequent nonlinear coupled system of equations is solved by the Lax-Wendroff scheme and is then applied to an open model arterial network of the human vascular system containing the largest fifty-five arteries. The proposed model with functional coefficients is compared with current classical one-dimensional theories which assume steady state Hagen-Poiseuille velocity profiles, either parabolic or plug-like, throughout the whole arterial tree. The effects of the nonlinear term in the momentum equation and different strategies for bifurcation points in the network, as well as the various lumped parameter outflow boundary conditions for distal terminal points are also analyzed. The results show that the proposed model can be used as an efficient tool for investigating the dynamics of reduced-order models of flows in physiological systems and would, in particular, be a good candidate for the one-dimensional, system-level component of geometric multiscale models of physiological systems.
\end{abstract}



\section{Introduction}
\label{sec:intro}
Extensive research has been done on modeling human physiology. Most of the work has been aimed at developing detailed models of specific components of physiological systems such as a cell, vein, molecule, or heart valve. One example of such an effort is the three-dimensional (3D) computational model of the heart developed at NYU by Peskin and McQueen \cite{peskin1989three}. The challenges of the interaction of mechanics with biology and medicine have recently been outlined by Bugliarello \cite{bugliarello2009mechanics}. In particular, modeling of internal flows through elastic vessels has also been studied intensively over years \cite{ku1997blood,grotberg2004biofluid,pedley2003mathematical,taylor2004experimental,humphrey2009vascular}, and a brief history of the arterial fluid mechanics can be found in a recent paper by Parker \cite{parker2009brief}. Wide range of the proposed models may be classified by their dimensionality. As far as those models are concerned, depending on the phenomena to be studied, numerous degrees of simplifications can be considered going from fully 3D fluid-structure interaction models \cite{verdonck2000intra,guidoboni2009kinematically,varshney2010numerical} to the simplified reduced-order models \cite{hughes1973one,ottesen2004applied,taylor2009patient}, and from complex non-Newtonian fluid models \cite{kafka2009constitutive,akbar2010simulation,nadeem2010peristaltic,nadeem2010simulation,nadeem2010series,nadeem2011inf67} to idealized Newtonian fluid models \cite{nichols2005mcdonald,chakravarty2009mathematical,grinberg2009large}.

There has been recent interest in large-scale fully 3D computations of patient-specific human arterial tree because of their importance in the emerging field of personalized medicine. This 3D numerical analysis, which is generally based on MRI imaging of the patient's vasculature, provides a very detailed view of local flow features such as recirculation and shear stress in a patient's circulatory system, but is enormously computationally intensive. Grinberg et al. report in \cite{grinberg2009large} that simulating one cardiac cycle in an arterial tree that included the largest arteries required 27.7 hours of computational time using 40,000 processors. While fully 3D simulations such as these are invaluable to our understanding of human physiology, and to treating common circulatory system disorders such as arterial stenosis and heart valve malfunction, most hospitals and clinics do not have ready access to supercomputing clusters, and an approach that is still very accurate, but less computationally intensive is desired in the clinical or hospital setting.  Global computational models of patients' entire cardiac functioning that can give useful results in a reasonable amount of time on a desktop computer are needed. Some of geometric multiscale models of the cardiovascular system have been recently developed that join fully detailed 3D simulations of a region of interest in a patient, for example, the carotid artery for a carotid artery stenosis, with global, macroscopic level reduced-order models for the rest of the patient's cardiovascular system, thereby reducing the spatial and computational complexity of the model \cite{formaggia1999multiscale,quarteroni2001modeling,passerini20093d,taylor2009patient,oran2009computational}.

The simplest reduced-order models for patient vasculature are spatially homogeneous, zeroth-order (0D) lumped parameter models based on an analogy between the cardiovascular system and an electrical circuit. In this approach, each segment of the system is modeled by basic circuit elements yielding a set of ordinary differential equations in the time domain. The complex network that comprises the human circulatory system can be simulated efficiently in this way \cite{westerhof1969analog,olufsen1999structured,olufsen2004deriving}. The main disadvantage of these 0D models is that they are unable to take into account some important features of cardiovascular function, such as wave propagation along vessels, which is patient-specific and critical to the accuracy of the detailed, microscopic level models.

Another approach to reduced-order modeling of the cardiovascular system is to develop spatially one-dimensional (1D) linear or nonlinear models consisting of coupled partial differential equations that describe the behavior of the venous cross sectional area, the averaged blood velocity, and the averaged blood pressure throughout the system \cite{taylor2004experimental,quarteroni2000computational,pedley2003mathematical,van2003mathematical,sherwin2003one,hughes1973one,stergiopulos1992computer,sheng1995computational,wan2002one,wang2004wave,vignon2004outflow,formaggia2003one,matthys2007pulse,alastruey2009analysing,marchandise2009numerical,liang2009multi}. These reduced-order 1D modeling efforts offers an excellent compromise with anatomic accuracy and has a great potential in clinical and research applications \cite{taylor2004experimental}, as well as in the other mechanical systems such as valveless impedance pumping for open or closed configurations \cite{timmermann2009novel}. They are based on the assumption that the flow velocity along the elastic vessels is much greater than the flow velocity perpendicular to the longitudinal axis, resulting a nonlinear coupled system of averaged continuity and momentum equations. An additional equation is used to close the system by providing the constitutive relationship between the cross-sectional area at a point along the system, and the pressure at that point.  In the geometric multiscale approach to modeling the cardiovascular system, these 1D models of a patient's vasculature are often truncated after a number of arterial tree generations, and supplemented with appropriate boundary conditions provided by 0D lumped parameter models that model the effects of the rest (the distal portion) of the vasculature \cite{olufsen1999structured,olufsen2004deriving,logana2005multiscale,grinberg2008outflow}.

As the 1D modeling and computations have become more or less routine, attention has turned to improvement the basic theory, especially in the context of bifurcations \cite{sherwin2003one}, elastic wall modeling \cite{payne2007methods,oscuii2007biomechanical}, and permanent constrictions (i.e, stenoses \cite{wan2002one}) or forced constrictions (i.e., valveless pumping \cite{timmermann2009novel}). In the 1D theory, momentum equation is derived by integrating an assumed velocity profile over the cross-sectional area of the tube. In all 1D modeling studies cited above, the assumed velocity profile has either plug-like form (corresponding to large Womersley numbers for larger vessels) or parabolic form (corresponding to small Womersley numbers for smaller vessels). In this classical formulation, the effect of changing shape/frequency parameter (Womersley numbers) on pulsatile flow is neglected due to the approximated steady state Hagen–-Poiseuille velocity profile giving a momentum equation with constant coefficients. In this study, a new one-dimensional model is derived by integrating the momentum equation over the transient Womersley flow velocity profile. Therefore, the coefficients in the averaged momentum equation changes continuously via Womersley numbers allowing us to simulate global physiological systems more appropriately.

Many numerical methods have been proposed to solve hyperbolic system of equations can be used for 1D modeling, for example, method of characteristics \cite{schaaf1972digital}, finite difference \cite{stergiopulos1992computer,sheng1995computational} and finite/spectral element methods \cite{sherwin2003one,marchandise2009numerical}. The two-step Lax-Wendroff scheme in the finite difference methods and its finite element equivalent form, Taylor-Galerkin scheme are common examples of second-order accurate standard algorithms for computing these flows. The finite-difference discretization with Lax-Wendroff scheme is used in this study to solve the improved reduced-order pulsed flow equations. It is second-order accurate in time and space. The PFE can accommodate multiple branching, nonlinear elastic wall effects, tapering vessels, some two-dimensional viscous and inertial effects, and different boundary conditions, as well as the effects of external forces such as gravity and outperforms the classical one-dimensional theory for frequecy/shape parameter changing internal flows such as flows in arterial tree.

The outline of this paper is as follows. In section 2, we first introduce the Pulsed Flow Equations (PFE) by extending the classical 1D theory with assumed Womersley velocity profiles changing continuously via a shape parameter. The different strategies for boundary conditions at terminal and bifurcation points are also presented in this section. The Lax-Wendroff scheme is introduced in section 3 in order to solve PFE efficiently. In section 4, the model and algorithm is then applied to an open model of a typical arterial network containing the largest fifty-five arteries with proximal boundary conditions at the Ascending Aorta, coupled with a lumped parameter model at the distal points of the vasculature to provide outflow boundary conditions. The relative importance of modeling strategies such as the influence of the assumed velocity profiles in 1D modeling, various models for bifurcation and terminal point boundary conditions is also analyzed and compared each others. Finally, in section 5, we present some conclusions and perspectives related to this work.

\section{Mathematical Modeling}
\label{sec:math}
\subsection{Derivation of the Pulsed Flow Equations}
\label{sec:math}
The mathematical model is based on the mass and momentum conservation equations of incompressible fluids streaming through the tubes, assuming axisymmetric flow of constant fluid density, temperature, and viscosity. The continuity and axial momentum equations of incompressible Newtonian fluid (a valid assumptions for blood flows through the large and medium vessels \cite{hughes1973one}) are given in axisymmetric cylindrical coordinates as:
\begin{equation}\label{eq:}
\frac{\partial (rv)}{\partial r} +\frac{\partial (ru)}{\partial x} = 0
\end{equation}
\begin{equation}\label{eq:2Dmom}
\frac{\partial u}{\partial t} +u\frac{\partial u}{\partial x} +v\frac{\partial u}{\partial r} =-\frac{1}{\rho}\frac{\partial p}{\partial x} +\nu(\frac{1}{r}\frac{\partial}{\partial r}(r\frac{\partial u}{\partial r}) + \frac{\partial^2 u}{\partial x^2})
\end{equation}
where $r$ and $x$ are the coordinates in radial and axial directions. Here, $\rho$ is the density, $\nu$ is the kinematic viscosity, $p$ is the pressure, $u$ and $v$ are the velocity components in axial and radial directions. Since the blood density is constant and the flow in confined in the artery, the gravity effect can be included in the pressure term. We seek to derive a reduced-order system of equations that govern the averaged axial velocity, $\bar{u}$, the pressure, $p$, and the cross-sectional area, $A$, in internal biological fluid flows, while capturing some essential effects occurring in biological fluid flows that are not captured by the standard equations for incompressible fluid flows in one-dimension. This can be accomplished by integrating the equations over the cross-sectional area. Integrating continuity equation over the cross-section results in:
\begin{equation}\label{eq:avecon}
\frac{\partial A}{\partial t} + \frac{\partial (\bar{u}  A)}{\partial x} = 0
\end{equation}
where we used $v(R)= \frac{\partial R}{\partial t}$. Here the cross-sectional area, $A=\pi R^2$, and the averaged velocity, $\bar{u} = \frac{2}{R^2} \int_{0}^{R} u(r,x,t)rdr$, are functions of one-dimensional space variable $x$, and time variable $t$. In order to integrate the axial momentum equation over the cross-section, an assumed velocity profile need to be taken. It is assumed that the velocity profile has the form:
\begin{equation}\label{eq:appr}
u(r,x,t) = f(r) \bar{u}(x,t)
\end{equation}
In the classical one-dimensional models, $f(r)$ is taken as steady state Hagen–-Poiseuille (HP) profile:
\begin{equation}\label{eq:}
f^{\mbox{HP}}(r)=\frac{\gamma + 2}{\gamma} (1-(\frac{r}{R})^\gamma)
\end{equation}
where the coefficient $\gamma=2$ corresponds to fully developed, steady, incompressible, Newtonian flow in a cylindrical tube, which is also called Poiseuille flow, and is used in many studies \cite{hughes1973one,stergiopulos1992computer,sheng1995computational,wan2002one,wang2004wave,vignon2004outflow}. Measurements show that the parabolic velocity profile is not quite accurate for the larger arteries. Alternatively, the plug-like velocity profile with $\gamma=9$ is extensively applied for the arterial tree network \cite{formaggia2003one,matthys2007pulse,alastruey2009analysing,marchandise2009numerical,liang2009multi}. Here, instead of these approximations, we use the following profile corresponding to pulsatile Womersley flow\cite{zamir2000physics}:
\begin{equation}\label{eq:Wop}
f^{\mbox{Wo}}(r)=\Re\{\frac{\Lambda J_0(\Lambda) - \Lambda J_0(\frac{r}{R}\Lambda)}{\Lambda J_0(\Lambda) - 2J_1(\Lambda)} \}
\end{equation}
where $\Re$ represents the real part of the complex variable and $J_0$, and $J_1$ are the zeroth and first order Bessel functions of first kind \cite{abramowitz1964handbook}. Here $\Lambda$ is defined as:
\begin{equation}
\Lambda = (\frac{i-1}{\sqrt{2}}) \mbox{Wo}, \quad \mbox{Wo}=\sqrt{\frac{\omega}{\nu}}R(x,t)
\end{equation}
where $i^2=-1$, $\omega$ is the angular frequency of transient flow and $\mbox{Wo}$ is Womersley number which can be thought as frequency or shape parameter (representing the various size of tube radius). Integrating Eq.~\ref{eq:2Dmom} over the cross-section by using Eq.~\ref{eq:appr} with Eq.~\ref{eq:Wop}, the averaged momentum equation becomes:
\begin{equation}\label{eq:avemom}
\frac{\partial \bar{u} }{\partial t} + \alpha \bar{u} \frac{\partial \bar{u} }{\partial x} = -\frac{1}{\rho}\frac{\partial p}{\partial x} + \nu\frac{\partial^2 \bar{u}}{\partial x^2} -\beta \pi \frac{\bar{u}}{A}
\end{equation}
where the coefficients becomes:
\begin{equation}\label{eq:alpha}
\alpha = \Re\{ \chi\frac{\chi + \chi^*}{2} \}, \quad \chi = \sqrt{1+\frac{\Lambda^2[J_0^2(\Lambda) + J_1^2(\Lambda)] -4 J_1^2(\Lambda)}{[\Lambda J_0(\Lambda) - 2J_1(\Lambda)]^2}}
\end{equation}
\begin{equation}\label{eq:beta}
\beta = \Re\{ \frac{-2\Lambda^2 J_1(\Lambda)}{\Lambda J_0(\Lambda) - 2J_1(\Lambda)} \}
\end{equation}
where the asterisk represents the complex conjugate. In above formulation, we have used the identities:
\begin{equation}\label{eq:}
\frac{d J_0(r)}{dr}=-J_1(r), \quad \frac{d rJ_1(r)}{dr}= rJ_0(r)
\end{equation}
\begin{equation}\label{eq:}
\int_{0}^{R} J_0(\frac{r}{R}\Lambda) r dr = \frac{R^2}{\Lambda}J_1(\Lambda)
\end{equation}
\begin{equation}\label{eq:}
\int_{0}^{R} J_0^2(\frac{r}{R}\Lambda) r dr = \frac{R^2}{2}[J_0^2(\Lambda) + J_1^2(\Lambda)]
\end{equation}
\begin{equation}\label{eq:}
\int_{0}^{R} J_1(\frac{r}{R}\Lambda) dr = \frac{R}{\Lambda}[1-J_0(\Lambda)]
\end{equation}
On the other hand, by assuming Hagen–-Poiseuille velocity profiles the the coefficients of momentum equation, Eq.~\ref{eq:avemom}, are obtained as:
\begin{equation}\label{eq:}
\alpha = \frac{\gamma+2}{\gamma+1}, \quad \beta = 2(\gamma+2)
\end{equation}
which are constants either $\gamma=2$ for parabolic velocity profiles, or $\gamma=9$ plug-like velocity profiles. However, in our formulation given in Eqs.~\ref{eq:alpha}-\ref{eq:beta}, the coefficients of momentum equation become functions of Womersley number providing a continuous dynamical partial differential equation for flows through tubes with various sizes in diameter as occurred in the physiological systems. For example, corresponding Wo numbers for some arteries from largest to medium sizes in radius is shown in Table 1 where the blood is considered to have a constant dynamic viscosity of 0.0045 $Ns/m^{2}$, and a constant density of 1050 $kg/m^{3}$ with having 1 s oscillatory cycle period corresponding to 60 heart beats per minute.
\begin{table}[H]
\label{tab:1}       
\tbl{Womersley numbers for some vessels in the arterial tree network.}
{\begin{tabular}{@{}llllll@{}}
\toprule
Arteries & $L$ & $R_{top}$ & $R_{bot}$ & $c_{0}$ & $\mbox{Wo}$   \\
    & (cm) & (cm) & (cm) & (m/s) & ($T=1$s)  \\
\colrule
Ascending Aorta (1)     & 4.0   & 1.470  & 1.440 & 4.30 &  17.8 \\
Thoracic Aorta I (18)   & 5.2   & 0.999  & 0.999 & 4.30 &  12.1 \\
Abdominal Aorta I (28)  & 5.3   & 0.610  & 0.610 & 5.00 &  7.38 \\
Superior Mesenteric (34)& 5.9   & 0.435  & 0.435 & 5.10 &  5.27 \\
L. Carotid Artery (15)  & 20.8  & 0.370  & 0.370 & 5.30 &  4.48 \\
R. Femoral Artery (52)  & 44.3  & 0.259  & 0.190 & 8.60 &  3.14 \\
Celiac B (30)           & 1.0   & 0.200  & 0.200 & 7.30 &  2.42 \\
L. Interosseous  (24)   & 7.9   & 0.091  & 0.091 & 14.3 &  1.10 \\
\botrule
\end{tabular}}
\end{table}

\begin{figure}
\centering
\includegraphics[width=0.7\textwidth]{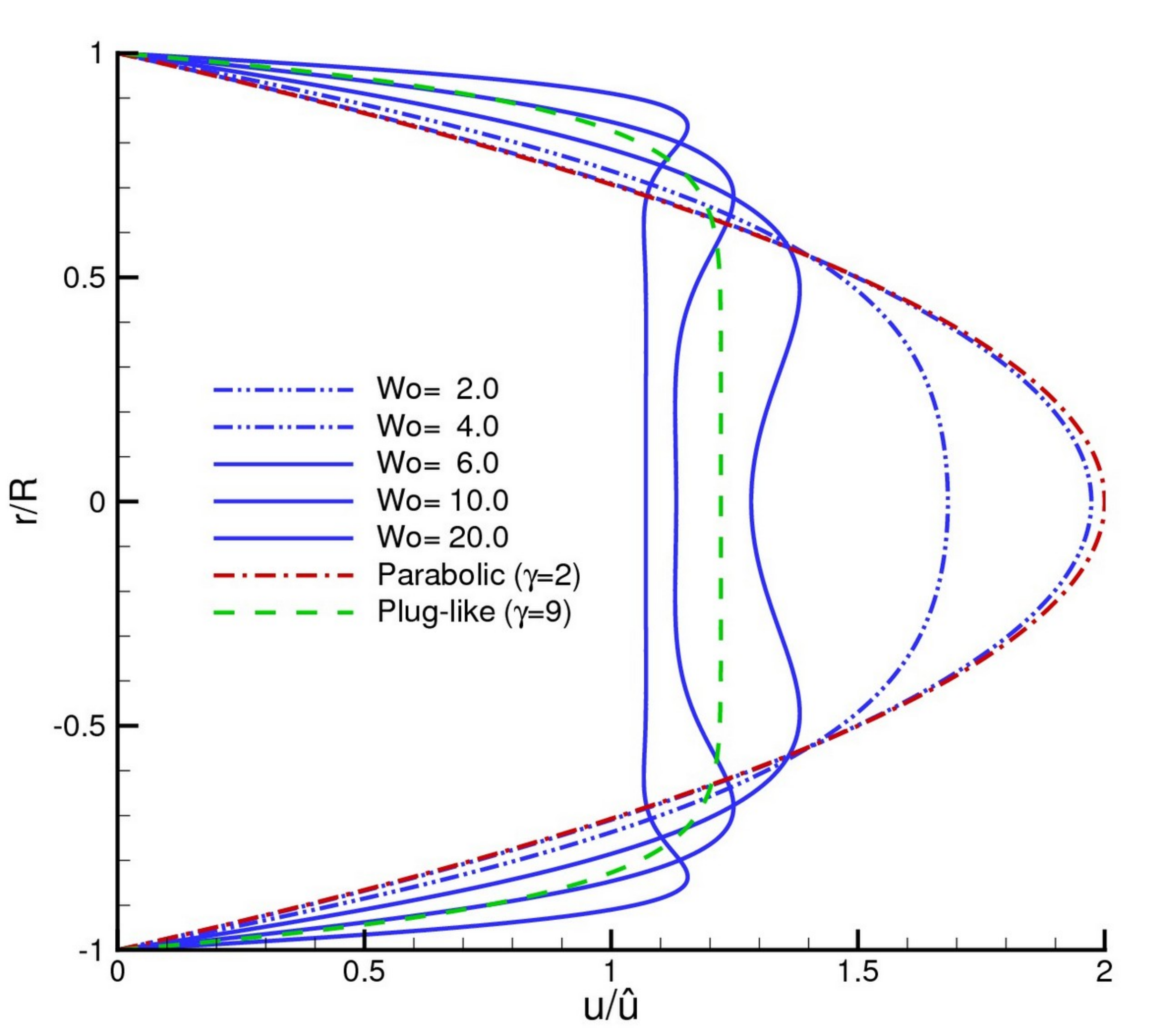}
\caption{Steady Hagen–-Poiseuille parabolic ($\gamma=2$) and plug-like($\gamma=9$) velocity profiles and five representative snapshots of the peak velocity profiles of the transient Womersley flow.}
\label{fig:profiles}
\end{figure}
\begin{figure}
\centering
\mbox{
\subfigure{\includegraphics[width=0.5\textwidth]{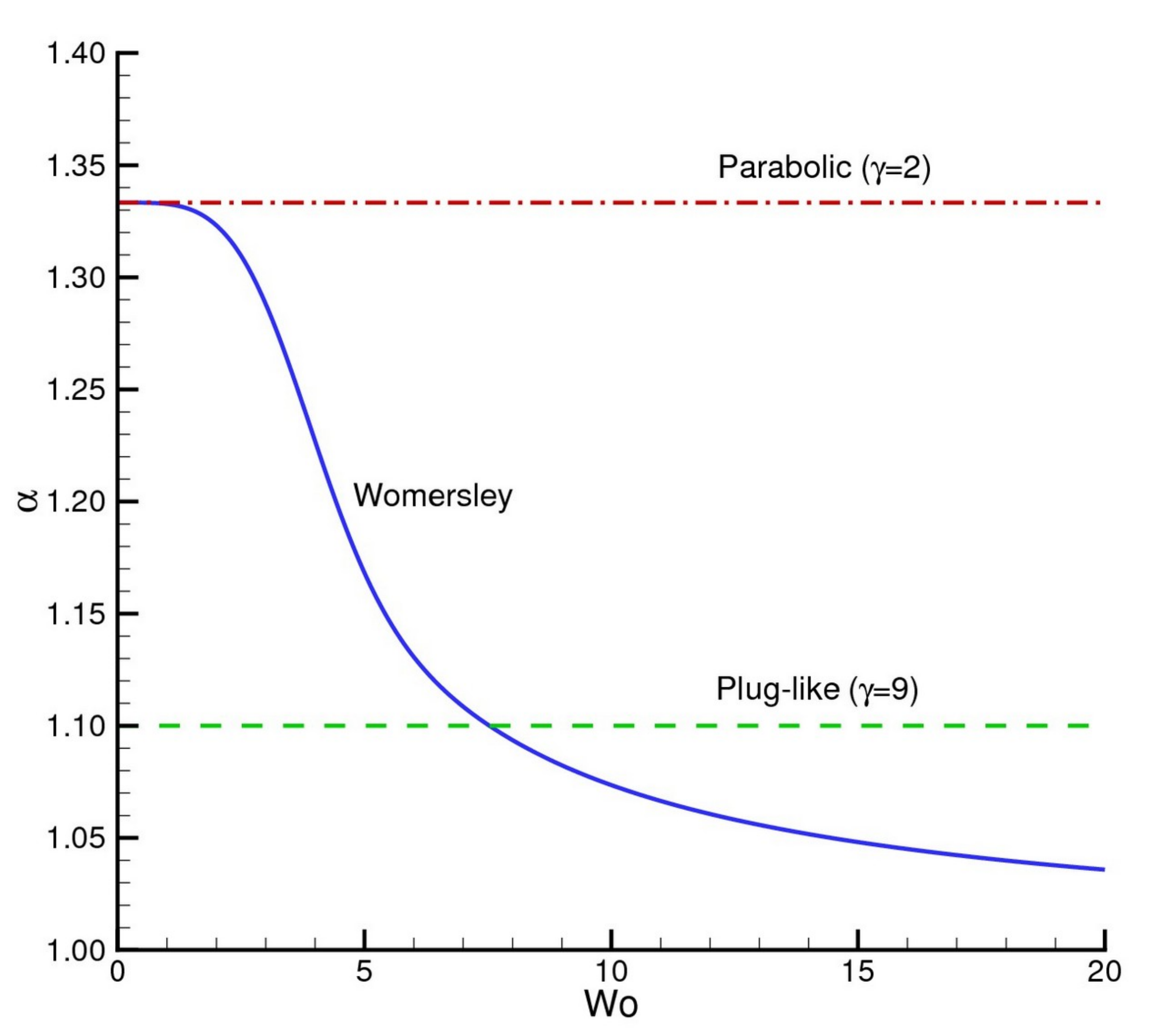}}
\subfigure{\includegraphics[width=0.5\textwidth]{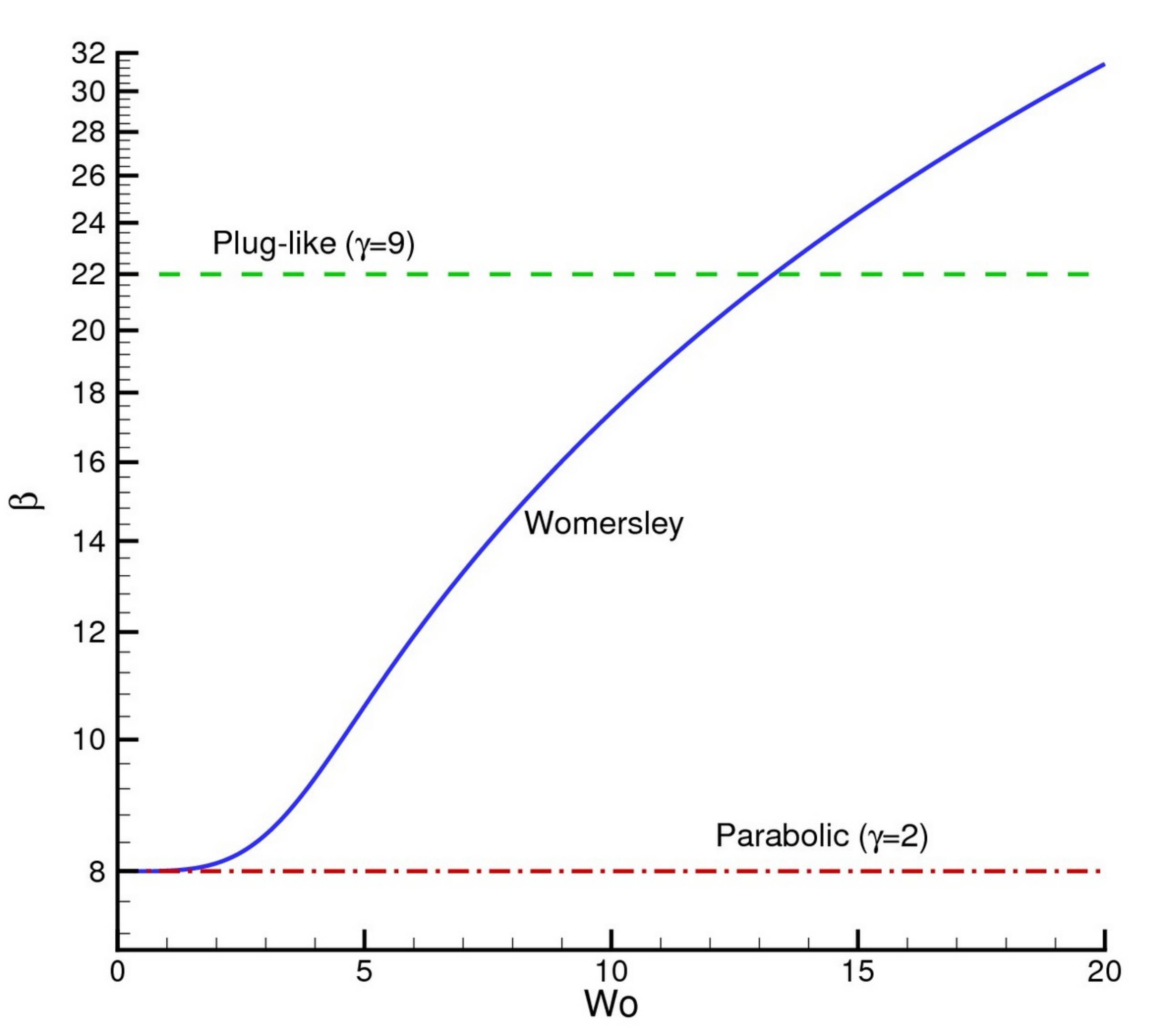}} }
\caption{Varying coefficients of (a) nonlinear term and (b) viscous drag term in the momentum equation.}
\label{fig:coeff}
\end{figure}
The normalized Womersley velocity profiles $u(r,x,t)/ \bar{u}(x,t) = f(r)$ are plotted in Fig.~\ref{fig:profiles}. The Hagen–-Poiseuille velocity profiles are also shown in the figure. For smaller Wo number, (Wo $ \leq $ 1), there is no significant difference between the Womersley and Hagen–-Poiseuille parabolic profile. The Eq.~\ref{eq:alpha} and Eq.~\ref{eq:beta} for coefficients of nonlinear and viscous drag terms in the momentum equation with varying Womersley number can be also reduced to the coefficients for classical 1D theory ($\gamma=2$) for small Womersley numbers. For convenience if we define, $\epsilon = \mbox{Wo}/\sqrt{2}$, then $\Lambda=\epsilon(i-1)$. By using the asymptotic formulas for Bessel functions with small arguments \cite{abramowitz1964handbook}, we have
\begin{equation}\label{eq:J0}
J_0(\Lambda) \sim 1 + i \frac{\epsilon^2}{2}
\end{equation}
\begin{equation}\label{eq:J1}
J_1(\Lambda) \sim \frac{\epsilon}{2}[-(1+\frac{\epsilon^2}{4})+i(1-\frac{\epsilon^2}{4})]
\end{equation}
Substituting Eqs.~\ref{eq:J0}-\ref{eq:J1} into the Eqs.~\ref{eq:alpha}-\ref{eq:beta}, we get the coefficients of classical theory:
\begin{equation}\label{eq:}
\alpha \sim \frac{4}{3},  \quad \beta \sim 8
\end{equation}
Since in the arterial tree Womersley number is bigger than that, therefore the classical theory has been extended for the plug-like velocity profile ($\gamma=9$) as mentioned before. This assumption approximates the velocity profile better for larger arteries, but takes one constant value all larger vessels. We improved one-dimensional theory by using the Womersley velocity profiles with functional coefficients, $\alpha$ and $\beta$ given in Eqs.~\ref{eq:alpha}-\ref{eq:beta}, of momentum equation which are shown in Fig.~\ref{fig:coeff}. Since Womersley parameter is changing through the tube segments in the network, the proposed functional form of $\alpha$ and $\beta$ is changing according to the Eqs.~\ref{eq:alpha}-\ref{eq:beta} through the network. This improvement in the one-dimensional theory allows us more accurate simulation of network type flows as occurred in human arterial tree for varying vessel radius. As mentioned earlier, previous studies for flows in human arterial tree have been performed using constant values for $\alpha$ and $\beta$ through the whole network. The influence of velocity profiles in 1D modeling is analyzed in section 4.
\subsection{Constitutive relationship}
\label{sec:cons}
In order to close the system of equations, the constitutive relation between the pressure and the cross-sectional area must be specified at a point in the system. Since the viscoelastic effects are small within the physiological range of pressure, the linear theory of elasticity can be used \cite{hughes1973one,olufsen1999structured,payne2007methods,taylor2004experimental}. According to Hooke's Law for thin walled tubes, the tangential and axial stresses and strains are related through the following formulas \cite{flugge1990stresses}:
\begin{eqnarray}
\varepsilon_t=\frac{\sigma_t}{E}-\frac{\upsilon \sigma_s}{E} \\
\varepsilon_s=\frac{\sigma_s}{E}-\frac{\upsilon \sigma_t}{E}
\label{eq:Hooke}
\end{eqnarray}
where $E$ is the Young's modulus of tube material and $\upsilon$ is the Poisson ratio, which is equal to $0.5$ for incompressible materials. Considering that the elongation of the tube is zero ($\varepsilon_s=0$), then the tangential strain becomes:
\begin{equation}
\varepsilon_t=\frac{(1-\upsilon^2)\sigma_t}{E}
\label{eq:TS}
\end{equation}
For a thin walled tube, the tangential tensile stress acting on it is
\begin{equation}
\sigma_t = \frac{R}{h}(p-p_{0})
\label{eq:TTS}
\end{equation}
where $p_{0}$ is the system's reference equilibrium pressure ($p$ is the fluid pressure at inner surface, and $p_{0}$ is the pressure at outer surface), and $h$ is the tube wall thickness. By using the definition of tangential strain, $\varepsilon_t=(2 \pi R- 2 \pi R_{0})/(2 \pi R_{0})$, and comparing it with Eq.~(\ref{eq:TS}), an algebraic expression of the pressure versus tube cross-sectional area is obtained:
\begin{equation}
p=p_{0} + \frac{E h}{R_{0}(1-\upsilon^2)}(1-\sqrt{\frac{A_{0}}{A} })
\label{eq:PA}
\end{equation}
where $A_{0}$ is the reference equilibrium area of the tube (the area when the $p=p_{0}$). The Young's modulus of the thin-walled tube is related the wave propagation speed by the Moens-Korteweg equation \cite{nichols2005mcdonald}:
\begin{equation}
c_{0} = \sqrt{\frac{Eh}{2 \rho R_{0}(1-\upsilon^2)}}
\label{eq:M-K}
\end{equation}
where $c_0$ is the wave propagation speed which depends on the properties of the elastic tube. Then, the constitutive relationship between pressure and area can be rewritten in the form of:
\begin{equation}
p=p_{0} + 2\rho(c_{0})^2(1-\sqrt{\frac{A_{0}}{A} })
\label{eq:RPA}
\end{equation}
As shown in Table 1, the corresponding wave propagation speed is in the range from $4$ to $15$ m/s and increases by decreasing vessel radius due to higher Young's modulus of smaller vessels. To account tapering effect for some of the vessels, the relationship for the radius along an individual artery is often specified in the following exponential relationship \cite{olufsen1999structured,ottesen2004applied}:
\begin{equation}
R_{0}(x)=R_{top} e^{ -ln(\frac{R_{bottom}} {R_{top}})\frac{x}{L}  }
\label{eq:taper}
\end{equation}
where $R_{top}$ and $R_{bottom}$ are the reference proximal and distal radii and $L$ is the length of the individual segment. The proximal (top) and distal (bottom) radius values under some reference condition for each particular segments for the human arterial system are well-documented in the literature \cite{ottesen2004applied,stergiopulos1992computer,wang2004wave,alastruey2009analysing}. In this study, the reference equilibrium radius values along the tubes through the network are computed by using Eq.~\ref{eq:taper}. Then the local Womersley number and the corresponding $\alpha$ and $\beta$ coefficients through the whole network can be pre-computed.

\subsection{Boundary conditions}
\label{sec:BC}
In order to finalize mathematical model, an appropriate boundary and initial conditions should be specified to the problem. For pulsatile flow problems, initial conditions does not have effect on the final solution. The results should convergence a periodic state after a few dummy simulation of oscillation cycles. Usually the hyperbolic system of equations has a positive wave propagation velocity much greater than the actual velocity of flow ($c_{0}\gg \bar{u}$). This is also true for blood flow computations in the arterial network. Consequently, we only need to specify either the area (pressure) or velocity values at the boundaries. Since the area and pressure are related each other by an algebraic constitutive relationship, Eq.~\ref{eq:RPA}, it is enough to specify pressure boundary conditions to the problem. For the open system of network flows, as studied in many arterial tree computations, the following boundary conditions must be established: (i) An equation at the inlet to the network; (ii) an equation at the outflow from each of terminal points of the network; (iii) a model at the each of the bifurcations.
\subsubsection{Inflow boundary condition}
\label{sec:BCin}
As an inflow boundary condition, we enforce either a time dependent area function, $A_{in}(t)$, or a time dependent velocity, $\bar{u}_{in}(t)$. In most blood flow simulations the flow rate, $Q_{in}(t)=\bar{u}A$, or pressure, $p_{in}(t)$  waveform can be specified as an inflow boundary condition at the proximal end of the network (i.e., the point where the arterial system is closest to the heart). In this work, we consider that the flow rate, $Q_{in}(t)$, is given as inflow boundary condition. The other variables can be computed with the help of discrete version continuity equation:
\begin{equation}
A_{0}^{n+1}=A_{0}^{n}-\frac{\Delta t}{\Delta x}(\bar{u}_{1}^{n+1}A_{1}^{n+1}-Q_{0}^{n+1})
\label{eq:dcon}
\end{equation}
where subscripts refer to spatial index, and superscripts denotes the discrete time level. The subscript $0$ locates the inlet point and the subscript $1$ locates the nearest discrete point to the inlet. Then, the corresponding velocity condition becomes trivial ($\bar{u}_{in}=Q_{in}/A_{0}^{n+1}$) and the inlet pressure is determined from Eq.~\ref{eq:RPA}. We have used this methodology for all the cases presented here.
\subsubsection{Outflow boundary condition}
\label{sec:BCout}
\begin{figure}
\centering
\includegraphics[width=1.0\textwidth]{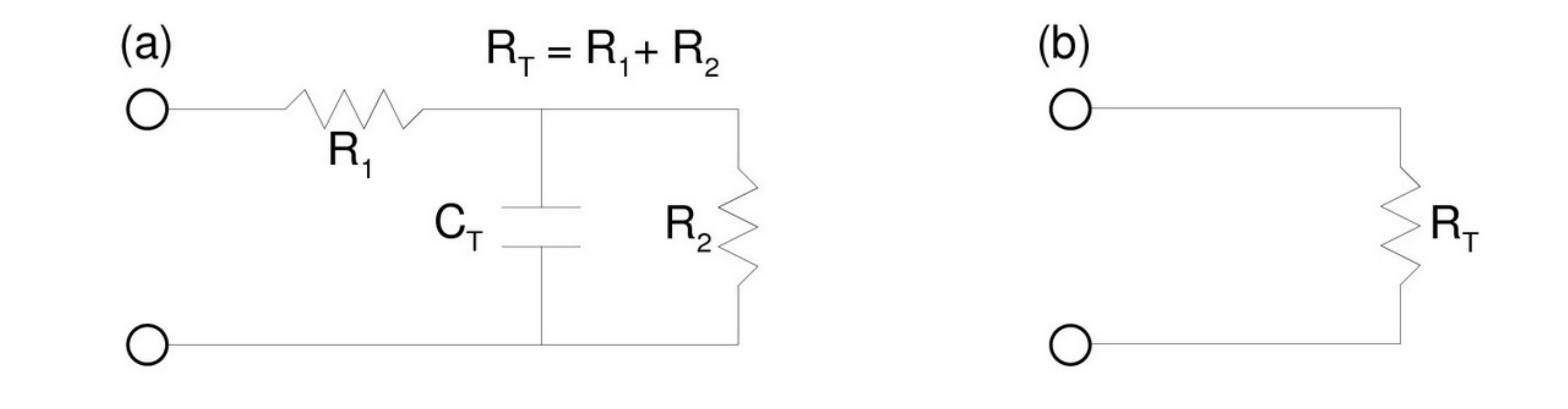}
\caption{Terminal (outflow) boundary conditions: (a) the three-element windkessel model with two resistor and one conductor, (b) pure resistor model.}
\label{fig:wk}
\end{figure}

For an open system arterial flow, the model is usually terminated using a lumped parameter impedance model for practical reasons. In the geometric multiscale approach to modeling the cardiovascular system, the 1D models of a patient's vasculature are often truncated after a number of arterial tree generations, and supplemented with appropriate boundary conditions provided by 0D lumped parameter models that model the effects of the rest (the distal portion) of the vasculature. The well accepted three-element windkessel model shown in Fig.~\ref{fig:wk}(a) with two resistors and a capacitor is used at the terminal points \cite{olufsen2004deriving,ottesen2004applied,stergiopulos1992computer}.
In the windkessel approach, the resistor elements represent the hydraulic resistance of the tube and the conductor element represents the elastic capacity of the tube. Using the circuit law, the corresponding relationship between the pressure and flow is given by:
\begin{equation}
\frac{\partial (\bar{u}A)}{\partial t}=\frac{1}{R_1}\frac{\partial p}{\partial t} + \frac{p}{R_1 R_2 C_T} - (1+\frac{R_1}{R_2})\frac{\bar{u}A}{R_1 C_T}
\label{eq:3WK}
\end{equation}
where $R_1+R_2=R_T$ is the total resistance and $C_T$ is the total compliance of the terminal branches. The major advantage of this model is that it accounts for both resistant and compliance effects of the distal vessels beyond the point of termination. Alternatively, a simpler pure resistance model shown in Fig.~\ref{fig:wk}(b) that does not consider compliance effects beyond the terminal points can also be used as an outflow boundary condition with the corresponding relationship given as:
\begin{equation}
\bar{u}=\frac{p}{R_T A}
\label{eq:wkres}
\end{equation}
These two lumped-parameter models are applied as outflow boundary conditions for terminal points and compared with each other in this study.
\subsubsection{Bifurcation point boundary conditions}
\label{sec:BCbif}
The one-dimensional theory derived for a flexible tube can be extended to handle the network flows by imposing a suitable interface condition at the branching points of the network. Assuming that all bifurcations occur at a point with one mother segment and two daughter segments as shown in Fig.~\ref{fig:bif}, the following nine quantities, $\bar{u}^{(1)}$, $A^{(1)}$, $p^{(1)}$, $\bar{u}^{(2)}$, $A^{(2)}$, $p^{(2)}$, $\bar{u}^{(3)}$, $A^{(3)}$, $p^{(3)}$, need to be specified at each bifurcation points. Since the pressure values are algebraically related to the area values via Eq.~\ref{eq:RPA}, there are only six independent unknown quantities for each bifurcation point in the network. The superscripts here indicate the segment numbers: (1) represents the mother tube, and (2) and (3) denotes the daughter tubes. Flow conditions at the bifurcation points can be represented by conservation of mass and pressure in the following form \cite{stergiopulos1992computer}:
\begin{figure}
\centering
\includegraphics[width=0.6\textwidth]{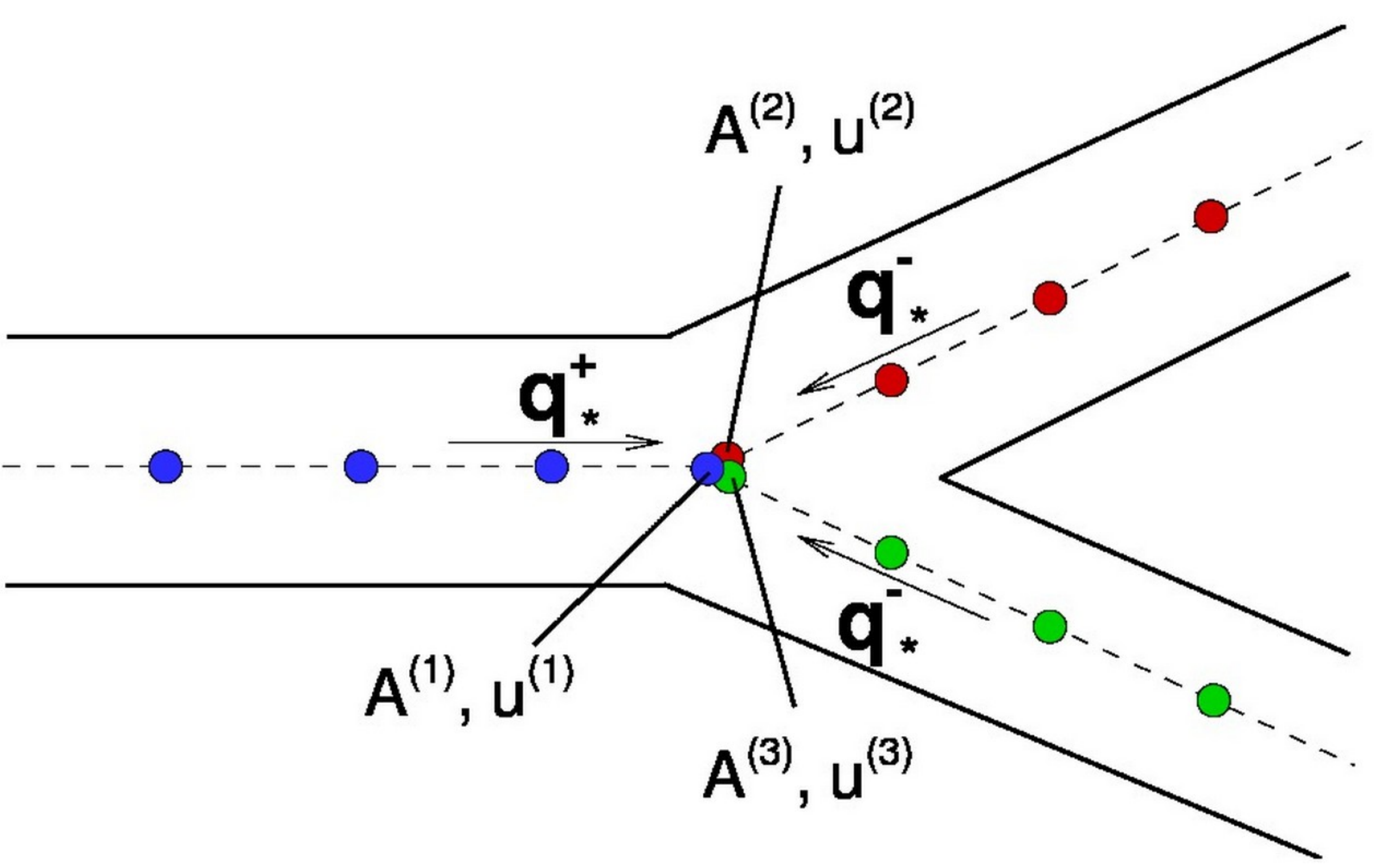}
\caption{The schematics for the bifurcation boundary conditions. Superscript 1 denotes the mother segment and superscripts 2 and 3 denote the two daughter segments. The subscript $*$ represents the values for the nearest solution points to the boundary points. }
\label{fig:bif}
\end{figure}
\begin{equation}
\bar{u}^{(1)}A^{(1)}= \bar{u}^{(2)}A^{(2)} + \bar{u}^{(3)}A^{(3)}
\label{eq:comass}
\end{equation}
\begin{equation}
p^{(1)}  =  p^{(2)}
\label{eq:pr1}
\end{equation}
\begin{equation}
p^{(1)} =  p^{(3)}
\label{eq:pr2}
\end{equation}
Alternatively, the Eq.~\ref{eq:pr1} and Eq.~\ref{eq:pr2} for pressure equalities can be improved by using the total pressure equalities as \cite{formaggia2003one}:
\begin{equation}
p^{(1)} + \frac{1}{2}\rho (\bar{u}^{(1)})^2 =  p^{(2)} + \frac{1}{2}\rho (\bar{u}^{(2)})^2
\label{eq:pre1}
\end{equation}
\begin{equation}
p^{(1)} + \frac{1}{2}\rho (\bar{u}^{(1)})^2 =  p^{(3)} + \frac{1}{2}\rho (\bar{u}^{(3)})^2
\label{eq:pre2}
\end{equation}
The equations above provide three equations for six unknowns for each bifurcating point. Three additional equations come from either Riemann invariants \cite{sheng1995computational,sherwin2003one} or discrete continuity equations \cite{liang2009multi}. Looking at the problem from a characteristics point of view (using Riemann invariants), information can only reach the bifurcation point from within mother tube by a forward-traveling wave. Similarly, within the daughter segments, information can only reach the bifurcation point by a backward-traveling wave. For PFE equations, Riemann invariants are approximated as:
\begin{equation}
q^{\pm} = \bar{u} \pm 4 c_{0} [1-(A_{0}/A)^{1/4}]
\label{eq:rieman}
\end{equation}
The derivation of Eq.~\ref{eq:rieman} is shown in Appendix. Using the positive Riemann invariant for the mother tube and the negative invariants for the daughter tubes, the three additional equations are:
\begin{equation}
\bar{u}^{(1)} + 4 c_{0}^{(1)} [1-(A_{0}^{(1)}/A^{(1)})^{1/4}] = q^{+}_{*}
\label{eq:cr1}
\end{equation}
\begin{equation}
\bar{u}^{(2)} - 4 c_{0}^{(2)} [1-(A_{0}^{(2)}/A^{(2)})^{1/4}] = q^{-}_{*}
\label{eq:cr2}
\end{equation}
\begin{equation}
\bar{u}^{(3)} - 4 c_{0}^{(3)} [1-(A_{0}^{(3)}/A^{(3)})^{1/4}] = q^{-}_{*}
\label{eq:cr3}
\end{equation}
where $q^{\pm}_{*}$ are values that can be easily computed at the nearest solution points in the discretized system of equations.

Alternatively, the discrete version of the continuity equations can be used as closure equations. For the mother segment, it is given as:
\begin{equation}
\frac{A^{(1)}-A^{(1)n}}{\Delta t} + \frac{\bar{u}^{(1)}A^{(1)}-\bar{u}^{(1)}_{*}A^{(1)}_{*}}{\Delta x} =0
\label{eq:dcon1}
\end{equation}
Similarly for the daughters:
\begin{equation}
\frac{A^{(2)}-A^{(2)n}}{\Delta t} + \frac{\bar{u}^{(2)}_{*}A^{(2)}_{*}-\bar{u}^{(2)}A^{(2)}}{\Delta x} =0
\label{eq:dcon2}
\end{equation}
\begin{equation}
\frac{A^{(3)}-A^{(3)n}}{\Delta t} + \frac{\bar{u}^{(3)}_{*}A^{(3)}_{*}-\bar{u}^{(3)}A^{(3)}}{\Delta x} =0
\label{eq:dcon3}
\end{equation}
where the subscript $*$ represents the nearest discrete point to the bifurcation point in the vessel. The superscript $n$ represents the previous time-step values. With combination of the strategies described above we can construct four different treatments for the bifurcation points. The summary is given in Table 2, and the effect of these procedures to the global model will be discussed in Section 4. The six unknowns are then solved with the Newton-Raphson algorithm \cite{oran2001numerical} using a nonlinear set of six equations. We use the values from the previous time step as an initial guess in the Newton-Raphson algorithm. The Newton-Raphson algorithm converges quadratically in a few iterations, usually one or two iterations per time step.
\begin{table}
\label{tab:bifst}       
\tbl{Strategies for bifurcation boundary points for computing six unknowns.}
{\begin{tabular}{@{}ll@{}}
\toprule
Procedure set & Set of equations   \\
\colrule
Characteristics \& total pressure   & Eq.~\ref{eq:cr1}, Eq.~\ref{eq:cr2}, Eq.~\ref{eq:cr3}, Eq.~\ref{eq:comass}, Eq.~\ref{eq:pr1}, Eq.~\ref{eq:pr2}  \\
Discreate continuity \& total pressure   & Eq.~\ref{eq:dcon1}, Eq.~\ref{eq:dcon2}, Eq.~\ref{eq:dcon3}, Eq.~\ref{eq:comass}, Eq.~\ref{eq:pr1}, Eq.~\ref{eq:pr2}  \\
Characteristics \& pressure   & Eq.~\ref{eq:cr1}, Eq.~\ref{eq:cr2}, Eq.~\ref{eq:cr3}, Eq.~\ref{eq:comass}, Eq.~\ref{eq:pre1}, Eq.~\ref{eq:pre2}  \\
Discreate continuity \& pressure   & Eq.~\ref{eq:dcon1}, Eq.~\ref{eq:dcon2}, Eq.~\ref{eq:dcon3}, Eq.~\ref{eq:comass}, Eq.~\ref{eq:pre1}, Eq.~\ref{eq:pre2} \\
\botrule
\end{tabular}}
\end{table}
\section{Lax-Wendroff Scheme}
\label{sec:NM}
As stated earlier, the PFE can be solved by classical numerical methods for hyperbolic partial differential equations such as the MacCormack or Lax-Wendroff finite difference schemes or Taylor-Galerkin formulation in the finite element framework. Governing PFE that describe the temporal and spatial evolution of pressure, velocity and cross-sectional area along the elastic tubes and can be rewritten in the following vector form:
\begin{equation}
\frac{\partial Q}{\partial t}+\frac{\partial F}{\partial x} = S
\label{eq:GE}
\end{equation}
where
\[
Q = \left[\begin{array}{c}
A \\ \bar{u} \end{array} \right],  F = \left[\begin{array}{c}
A\bar{u} \\ \frac{\alpha}{2}\bar{u}^2 +2(c_{0})^2(1-\sqrt{\frac{A_{0}}{A} }) \end{array} \right], S=\left[\begin{array}{c}0 \\ -\frac{1}{\rho}\frac{\partial p_0}{\partial x} + \nu\frac{\partial^2 \bar{u}}{\partial x^2} -\beta \pi \nu\frac{\bar{u}}{A}\end{array}\right]
. \]
The problem given in Eq.~\ref{eq:GE} can be solved by Lax-Wendroff scheme using the following two-step procedure \cite{oran2001numerical}:
\begin{equation}
Q^{n+\frac{1}{2}}_{i+\frac{1}{2}} = \frac{Q^{n}_{i} +Q^{n}_{i+1} }{2}-\frac{1}{2}\frac{\Delta t}{\Delta x}(F^{n}_{i+1} -F^{n}_{i}) +\frac{\Delta t}{4} (S^{n}_{i}+S^{n}_{i+1})
\label{eq:r9}
\end{equation}
\begin{equation}
Q^{n+1}_{i} = Q^{n}_{i}-\frac{\Delta t}{\Delta x}(F^{n+\frac{1}{2}}_{i+\frac{1}{2}} -F^{n+\frac{1}{2}}_{i-\frac{1}{2}}) +\frac{\Delta t}{2}(S^{n+\frac{1}{2}}_{i+\frac{1}{2}}+S^{n+\frac{1}{2}}_{i-\frac{1}{2}})
\label{eq:r9}
\end{equation}
where superscripts and subscripts show the time index and spatial index, respectively. Inside each tube, the $n+1$ time-step values are obtained from the $n$ time-step values using two subsequent steps. The inflow/outflow boundary conditions and the internal boundary conditions at the bifurcation points are provided in the previous section. This scheme is second-order accurate in space and time. The stability criterion for the Lax-Wendroff scheme is
\begin{equation}\label{eq:CFL}
\mbox{CFL} = \frac{\Delta t (\bar{u} + c_0)} {\Delta x}  \leq 1
\end{equation}
where $c_0$ is the speed of wave propagation defined earlier by the Moens-Korteweg relationship.
\section{Results for arterial network}
\begin{figure}[h!]
  \centering
  \includegraphics[width=0.4\textwidth]{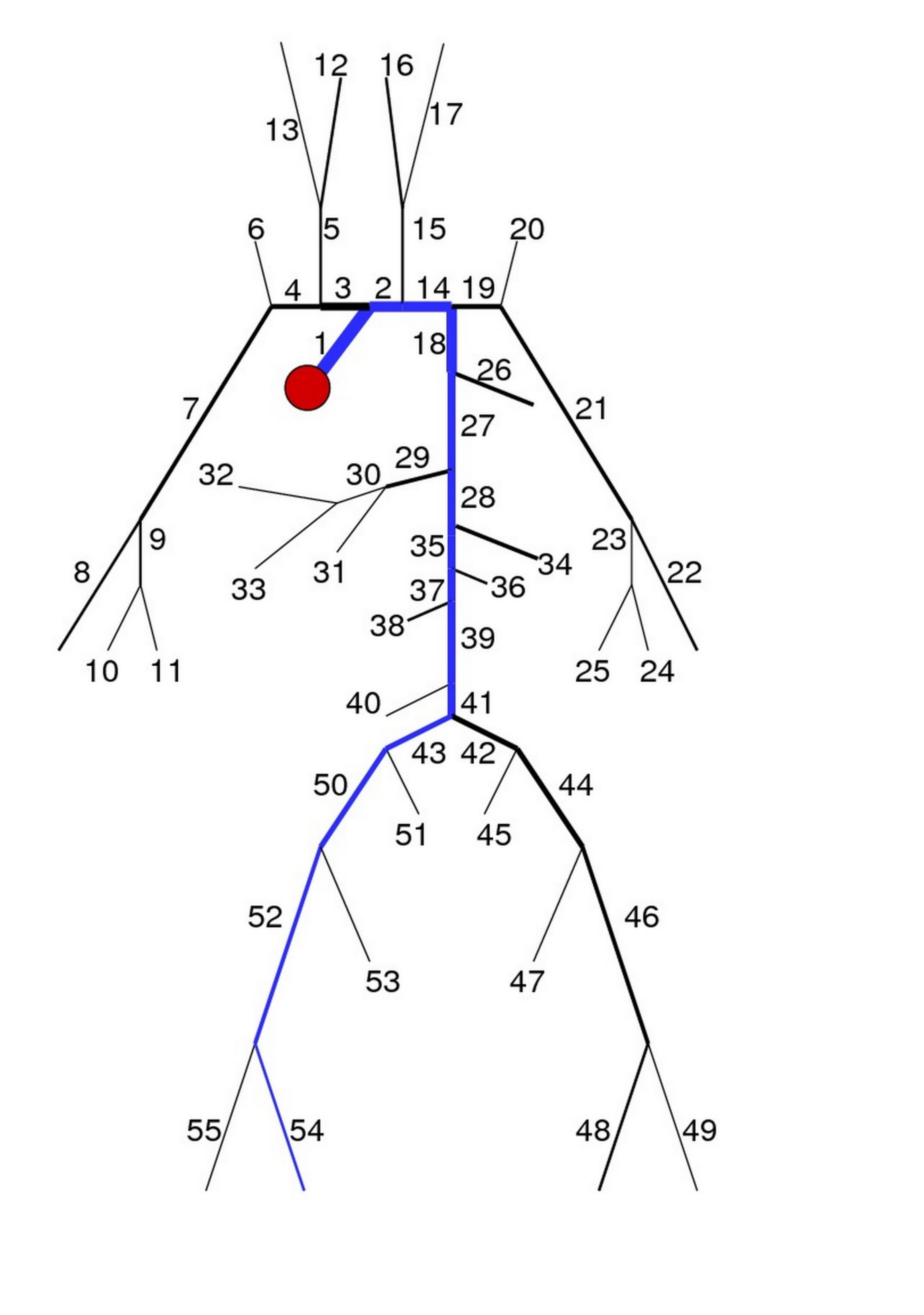}\\
  \caption{The arterial tree model including the largest fifty-five arteries.}
  \label{fig:tree}
\end{figure}

\begin{figure}[h!]
  \centering
  \includegraphics[width=0.6\textwidth]{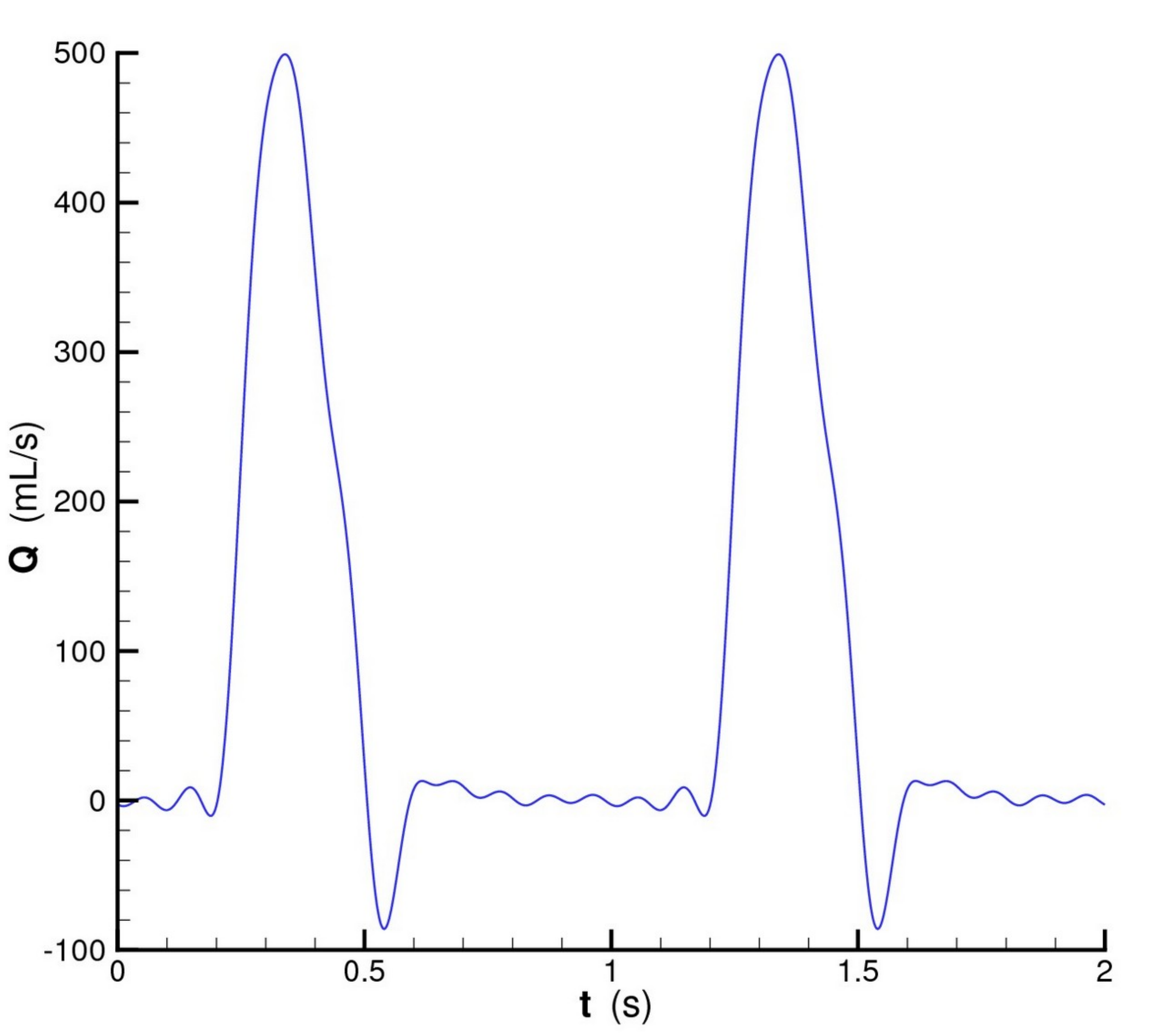}\\
  \caption{The periodic inflow for the proximal end of the Ascending Aorta (segment \# 1).}
  \label{fig:inflow}
\end{figure}

In this section a representative model for the human cardiovascular system containing the largest fifty-five arteries is simulated. The schematic description of this network flow is illustrated in Fig.~\ref{fig:tree}.
The physiological data including length, proximal and distal radius of every fifty-five segments and total resistance and compliance coefficients of terminal points and wave propagation speed (including the elastic parameters of the arteries) were prescribed based on the data reported in \cite{stergiopulos1992computer,wang2004wave,alastruey2009analysing}. The blood is considered to have a constant dynamic viscosity of 0.0045 $\mbox{Ns}/\mbox{m}^{2}$, and a constant density of 1050 $\mbox{kg}/\mbox{m}^{3}$. The equilibrium pressure is chosen as $P_0=98$ mmHg, which is assumed as the reference mean pressure corresponding to the equilibrium reference diameters of the vessels \cite{stergiopulos1992computer}. The flow waveform is specified at the proximal end of the network (at the beginning of the Ascending Aorta) by using the first ten Fourier harmonics \cite{stergiopulos1992computer} with the period of one second, which is plotted in Fig.~\ref{fig:inflow} corresponding to 60 heart beats per minute giving approximately 5.18 mL/min cardiac output to the arterial tree.

\begin{figure}
  \centering
  \includegraphics[width=1\textwidth]{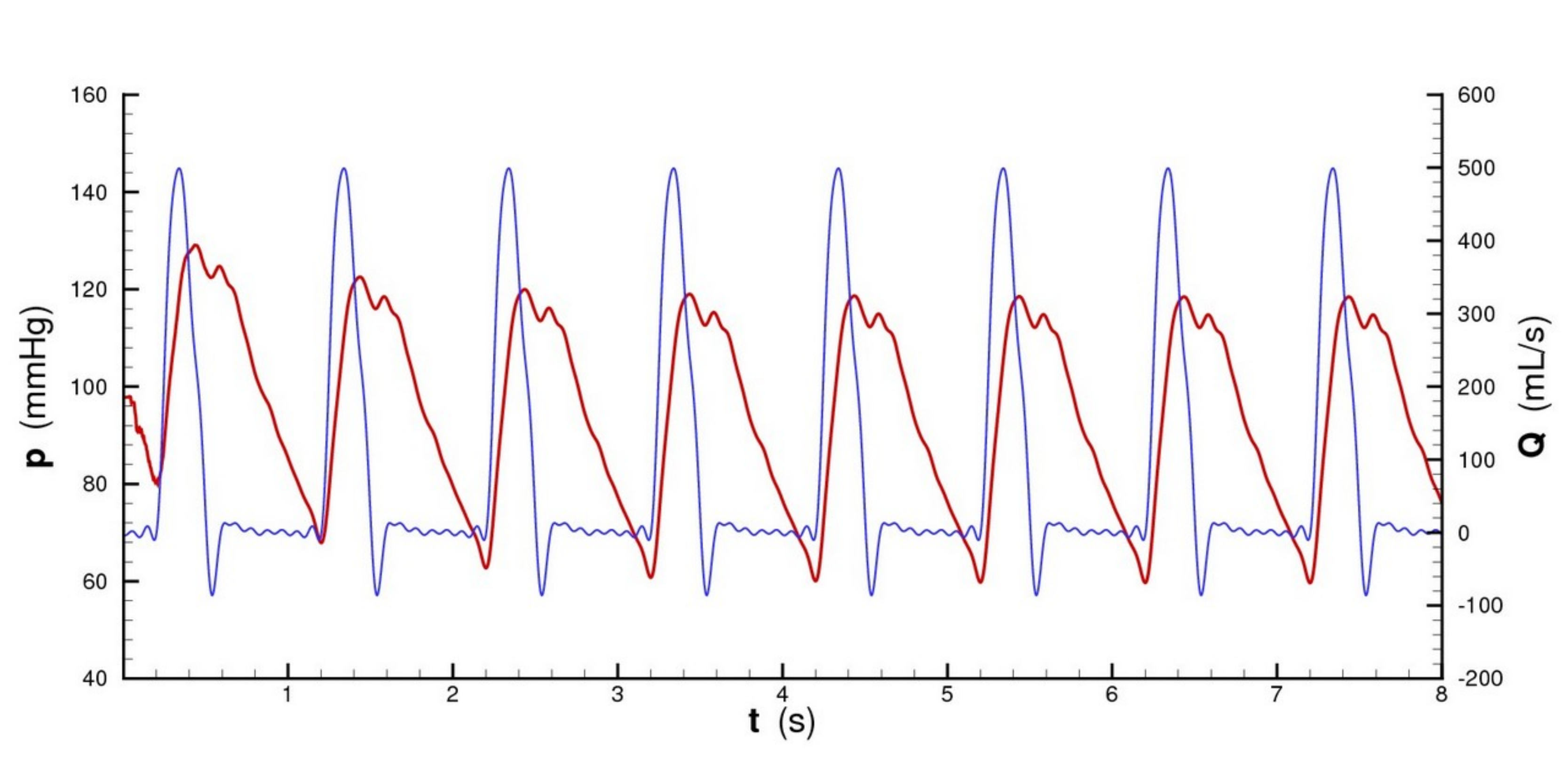}\\
  \caption{Convergence of pressure and flow waveforms at the inlet. The thin blue line shows the specified periodic inflow boundary condition, and the thick red line shows the pressure waveform.}
  \label{fig:inlet}
\end{figure}

Although this is a time dependent problem, exact initial conditions can not be imposed at all discrete points except at the proximal and distal boundary points explained in Section 2. Fortunately, the viscous effects damped out the discrepancies due to incorrect initial pressure and flow distribution, and in case of periodic flow the solution converges within two or three cycles as shown in the Fig.~\ref{fig:inlet}. The flow waveform is periodic (specified Fourier harmonics), and the pressure waveform at the inlet is converging after the third cycle. This convergence is true for all the segments in the network. Therefore we will present our converged results in rest of the paper.
\subsection{Grid dependance study}
\begin{figure}
\centering
\includegraphics[width=0.7\textwidth]{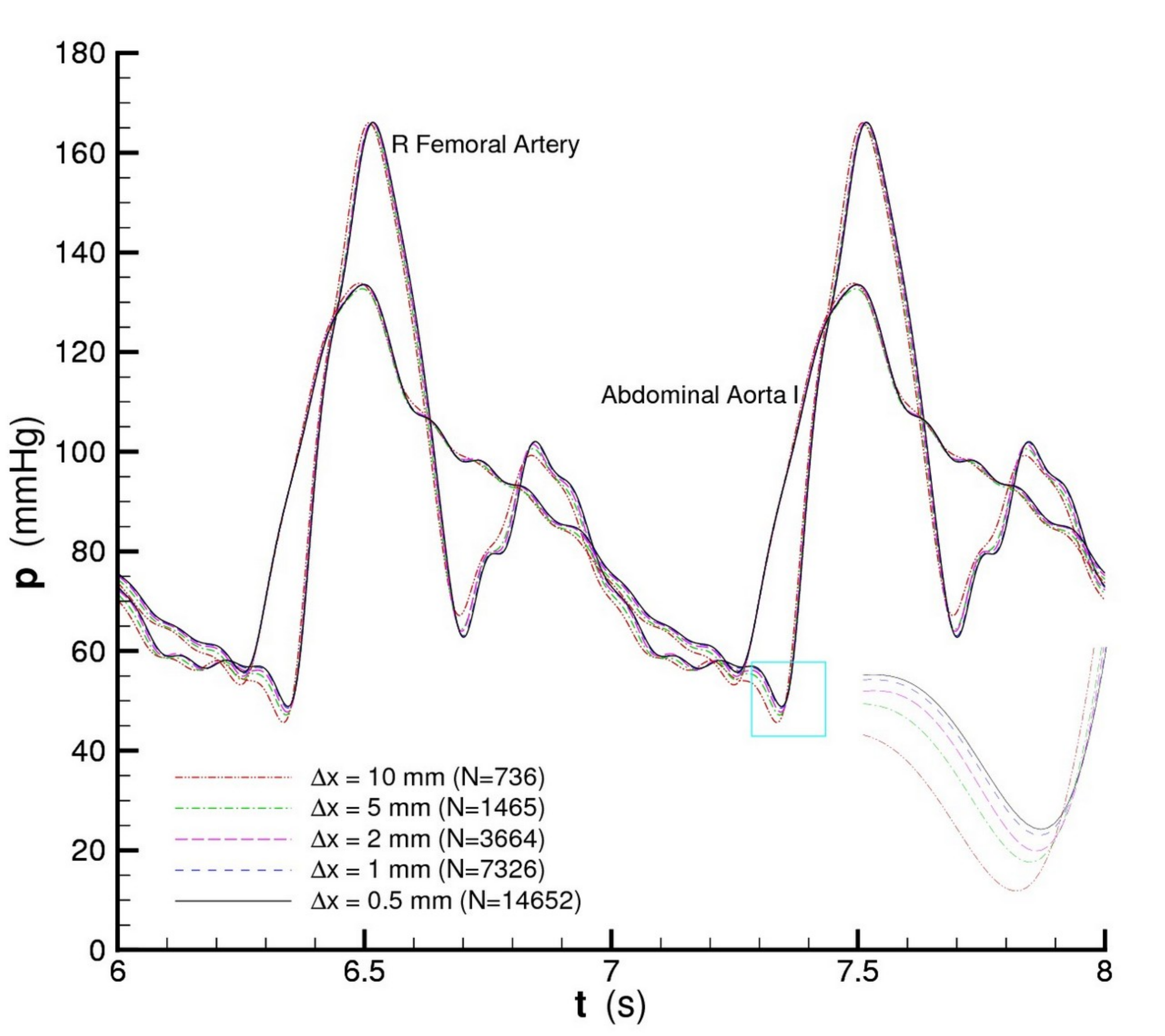}
\caption{Effects of varying grid resolutions on the pressure waveform.}
\label{fig:gridp}
\end{figure}

\begin{figure}
\centering
\includegraphics[width=0.7\textwidth]{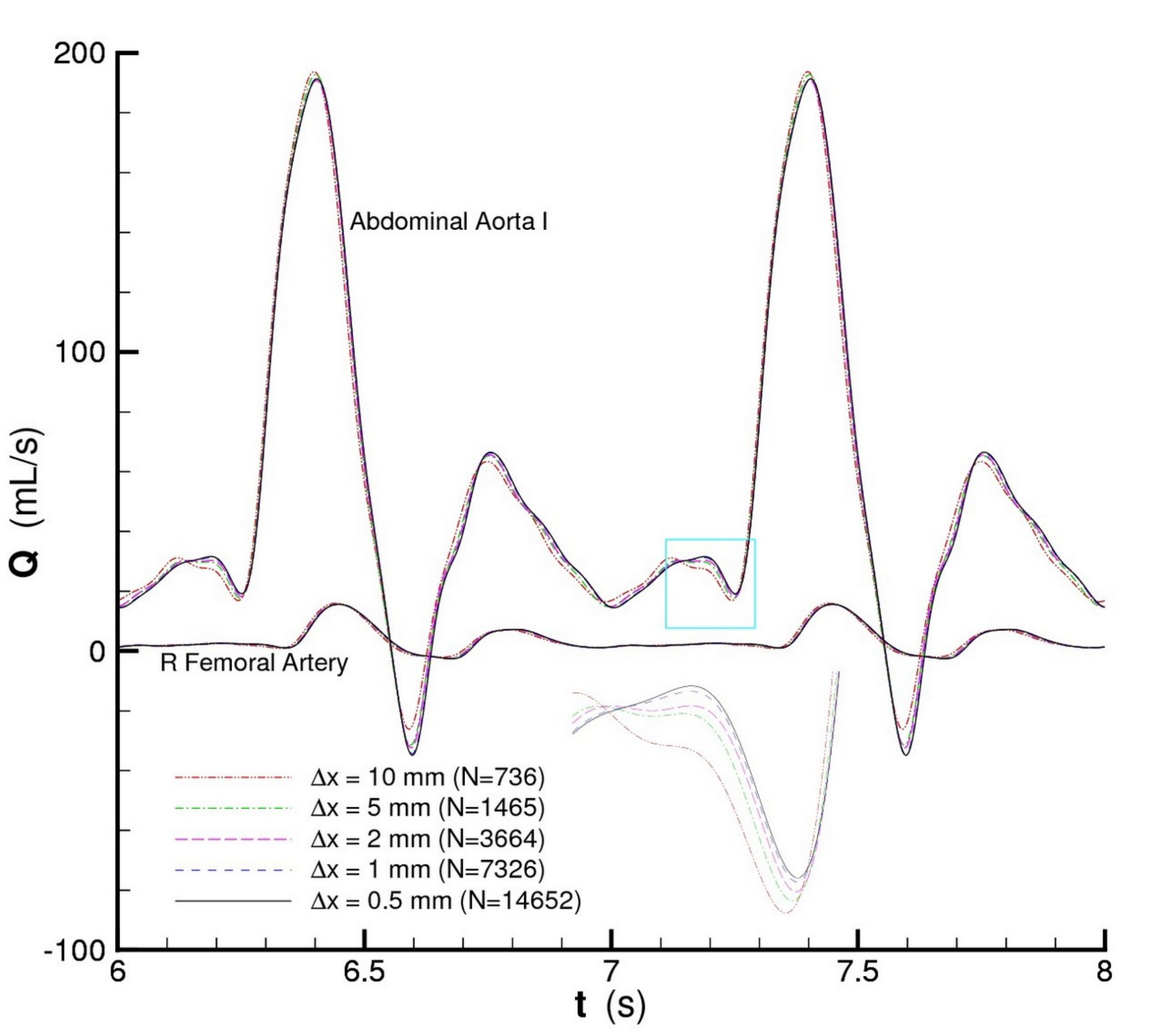}
\caption{Effects of varying grid resolutions on the flow waveform.}
\label{fig:gridq}
\end{figure}

We first investigate the effect of the the spatial resolution within arterial tree network with a small time step chosen as $\Delta t= 2.5\times 10^{-5}$ s. Comparisons of different resolutions on pressure and flow waveforms are shown in Fig.~\ref{fig:gridp} and Fig.~\ref{fig:gridq} for the distal point of the Abdominal Aorta I (segment \# 28), which is typical of a large artery, and for the distal point of the R. Femoral (segment \# 52), which is representative of a typical smaller sized artery. In figure labels, N represents the total number of grid points in the network. We can conclude that the $\Delta x = 2$ mm along the arterial tree is enough to capture all the waveform correctly. This was also outlined in \cite{sheng1995computational}. The computed waveforms are similar to those computed in \cite{stergiopulos1992computer}, and exhibit the general characteristics of the systemic circulations observed in experimentally \cite{nichols2005mcdonald}. The shapes of the pressure and flow waveforms were generally in good agrement with experimental measurement and the effects of the other mathematical modeling issues will be analyzed systemically in the following subsections.
\subsection{Effects of the nonlinear term in the momentum equation}
\begin{figure}
\centering
\includegraphics[width=0.7\textwidth]{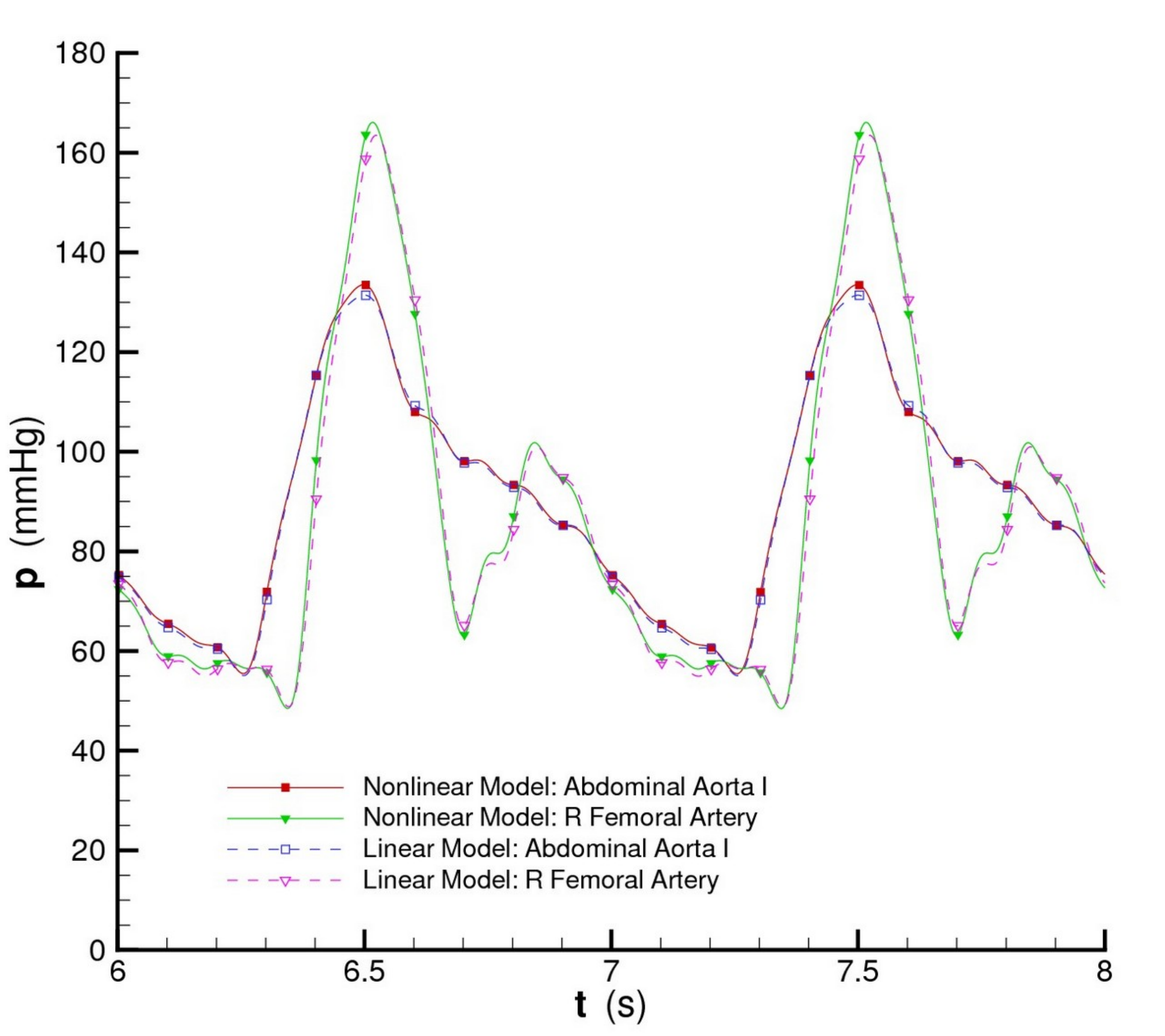}
\caption{Effects of the nonlinear term in the momentum equation on the pressure waveforms. Linear model assumes coefficient of the nonlinear term $\alpha=0$. }
\label{fig:nonp}
\end{figure}

\begin{figure}
\centering
\includegraphics[width=0.7\textwidth]{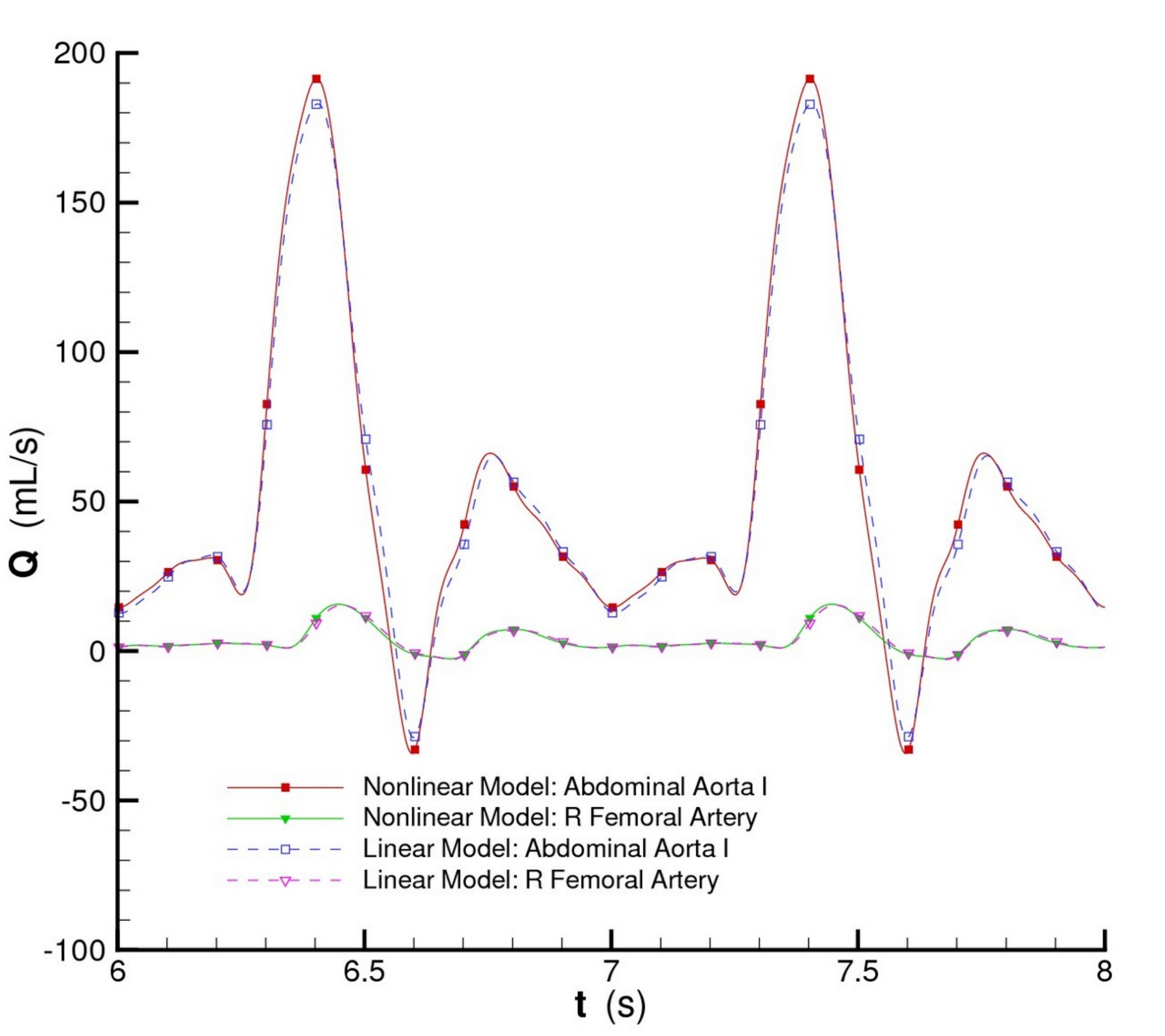}
\caption{Effect of the nonlinear term in the momentum equation on the flow waveforms. Linear model assumes coefficient of the nonlinear term $\alpha=0$.}
\label{fig:nonq}
\end{figure}

In order to quantify the effect of nonlinear term in the momentum equation, we perform two numerical runs such that one with $\alpha$ given in Eq.~\ref{eq:alpha} and one with $\alpha = 0$. Linear model assumes $\alpha=0$, and for nonlinear model, the $\alpha$ is changing through the arteries in the network and it is determined from Eq.~\ref{eq:alpha}. For both model, the other conditions and parameters are the same, i.e., the viscous drag term coefficient $\beta$ is computed from Eq.~\ref{eq:beta}. Computed pressure and flow waveforms are given in Fig.~\ref{fig:nonp} and Fig.~\ref{fig:nonq}, respectively. The model with nonlinear term estimates a little bit bigger pressure amplitude through the arterial tree. Since the wave propagation speed is much bigger than the actual velocity this small difference is expected. It can also be seen that there is very small phase shift. It should be noted that the momentum equation including nonlinear term will be used for rest of the simulations in this paper.

\subsection{Effects of the assumed velocity profiles}
\begin{figure}
\centering
\includegraphics[width=0.7\textwidth]{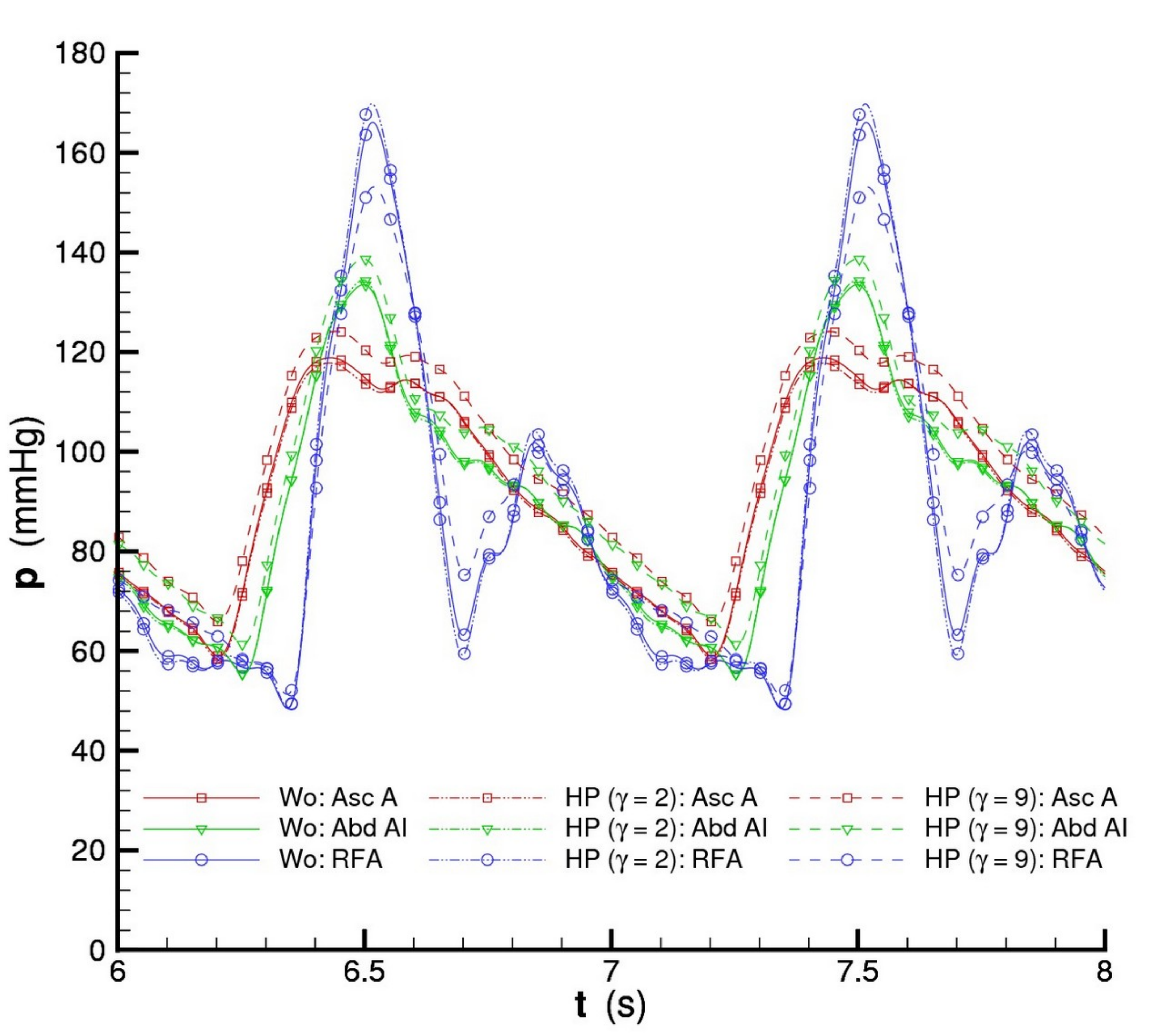}
\caption{Computed pressure waveforms with Hagen–-Poiseuille and Womersley velocity profiles at the Ascending Aorta (Asc A), Abdominal Aorta I (Abd AI) and Right Femoral Artery (RFA). HP($\gamma=2$) assumes the parabolic velocity profiles corresponding $\alpha=4/3$, and $\beta=8$; HP($\gamma=9$) assumes the plug-like velocity profiles corresponding $\alpha=11/10$, and $\beta=22$; and Wo assumes the functional coefficients (changing along the vessels in whole network) given by Eq.~\ref{eq:alpha} for $\alpha$ and Eq.~\ref{eq:beta} for $\beta$.}
\label{fig:model}
\end{figure}

\begin{figure}
\centering
\mbox{\subfigure{\includegraphics[width=0.35\textwidth,angle=270]{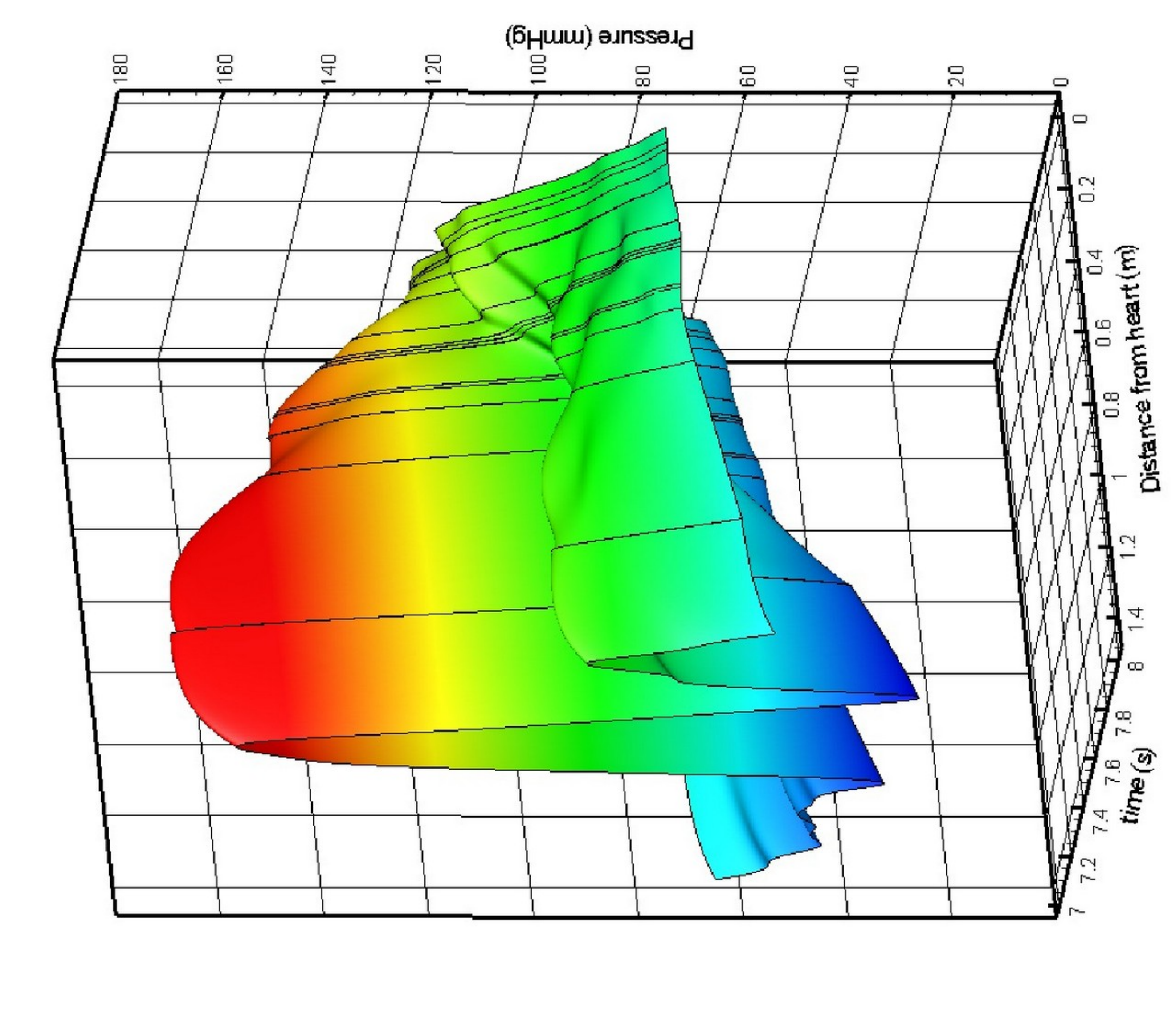}}
\subfigure{\includegraphics[width=0.35\textwidth,angle=270]{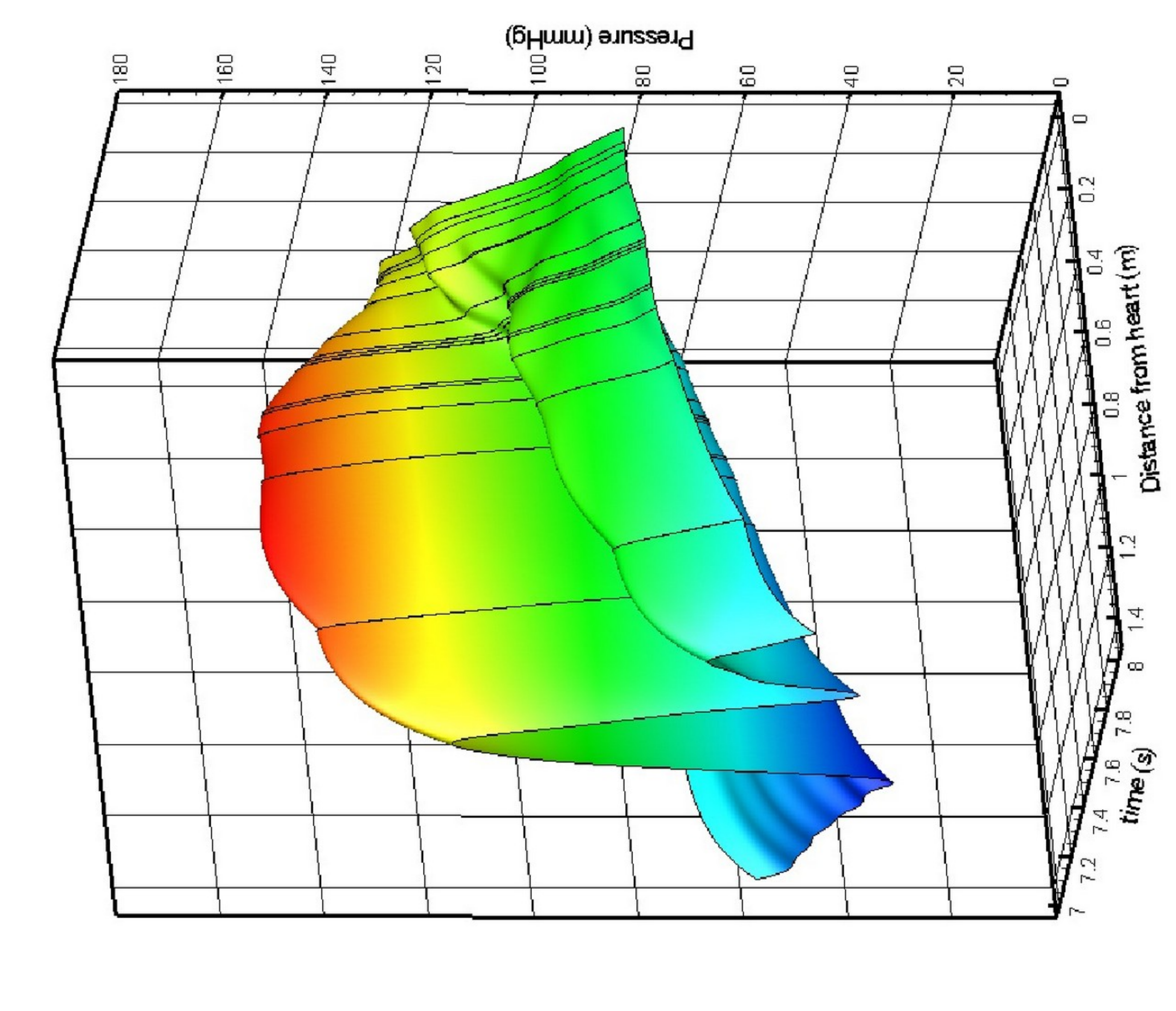} }
\subfigure{\includegraphics[width=0.35\textwidth,angle=270]{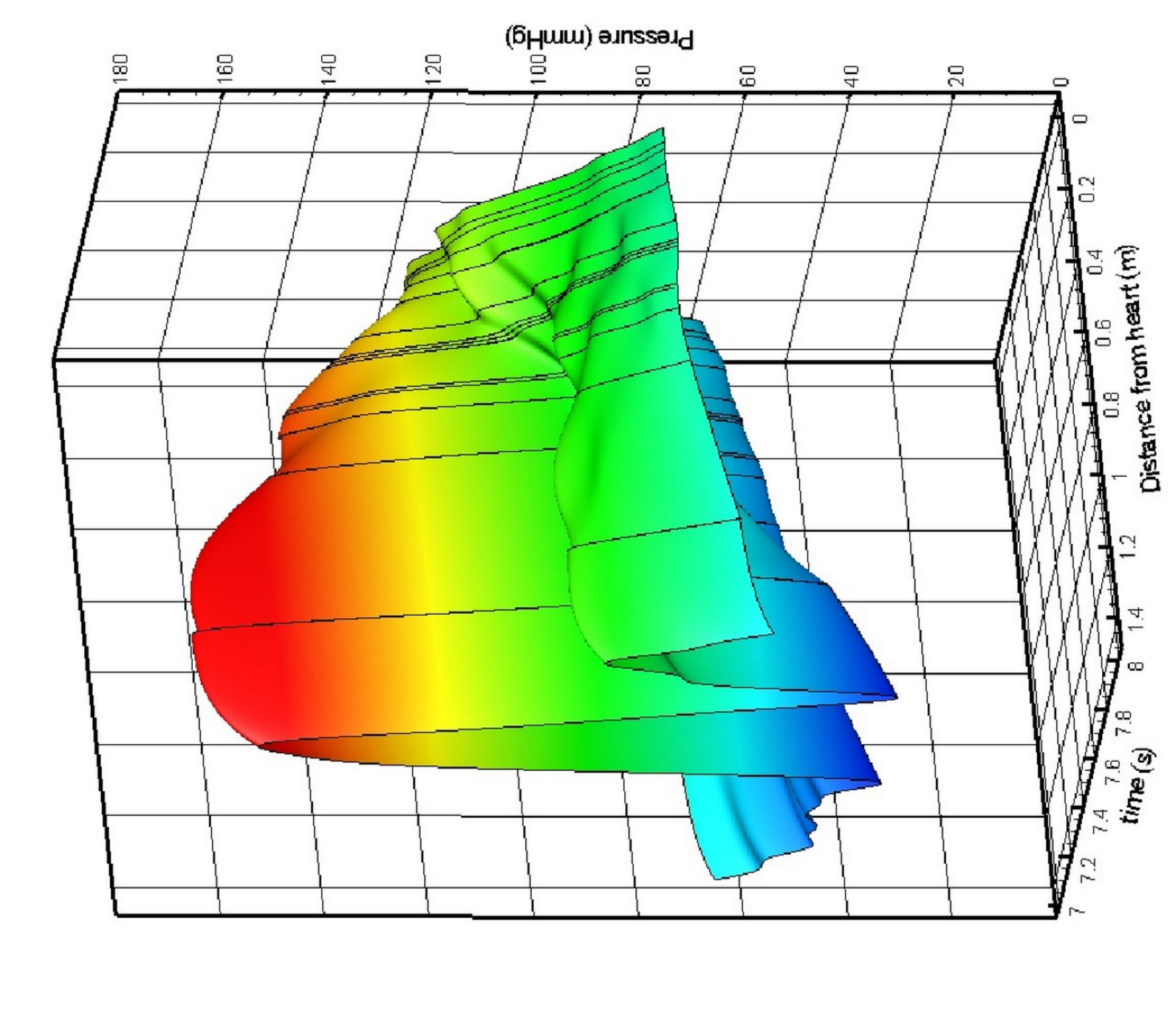} }}
\caption{Surface plots of the pressure waveforms along the Ascending Aorta up to the Right Posterior Tibial Artery (including 14 segments); (a) Hagen-Poiseuille parabolic velocity profile ($\beta=8$), (b) Hagen-Poiseuille plug-like velocity profile ($\beta=22$), (c) Womersley profile ($\beta$ from Eq.~\ref{eq:beta}) by varying $\beta$ values through the vessels in the network.}
\label{fig:3D}
\end{figure}

The coefficients of the momentum equation are based on the velocity profile that we assume when deriving pulsed flow equations. These coefficients are determined by the velocity profile over the cross-section of tube. Here we compare the classical one dimensional model with steady Hagen-Poiseuille parabolic velocity profile ($\gamma=2$), the plug-like velocity profile ($\gamma=9$), and the transient Womersley profile model proposed in this study. The effects of the assumed velocity profiles on the pressure waveforms are illustrated in Fig~\ref{fig:model} at three different segments: Ascending Aorta, Abdominal Aorta I, and Right Femoral Artery. It can be easily seen that the mathematical model with plug-like velocity profile overestimates the pressure waveform due to the higher viscous drag along the whole arterial tree. It should be noted that all the other modeling options such as bifurcation point treatment, boundary conditions, and grid size are chosen identical to emphasize the true effect of the assumed velocity profiles. In order to see the effect of the assumed velocity profiles along the whole arterial tree, in Fig.~\ref{fig:3D}, we perform corresponding surface pressure waveforms plot along the arteries form the Ascending Aorta to the Right Posterior Tibial Artery, colored blue in the Fig.~\ref{fig:tree}. It is clear that choosing of the velocity profile makes a large difference in arterial system modeling, and we present a global model with smoothly changing velocity profile along the arterial tree. However, if one would like to choose one of the Hagen-Poiseuille profiles, we conclude that the parabolic velocity profile seems a better choice to capture the waveform correctly instead of plug-like velocity profiles. Since the $\beta$ parameter are less than 22 in majority part of the arterial tree network, choosing plug-like velocity profiles (giving $\beta=22$) overestimates the pressure waveforms through the peripheral parts arterial tree.

\subsection{Effects of the bifurcation point modeling}
\begin{figure}
\centering
\includegraphics[width=0.7\textwidth]{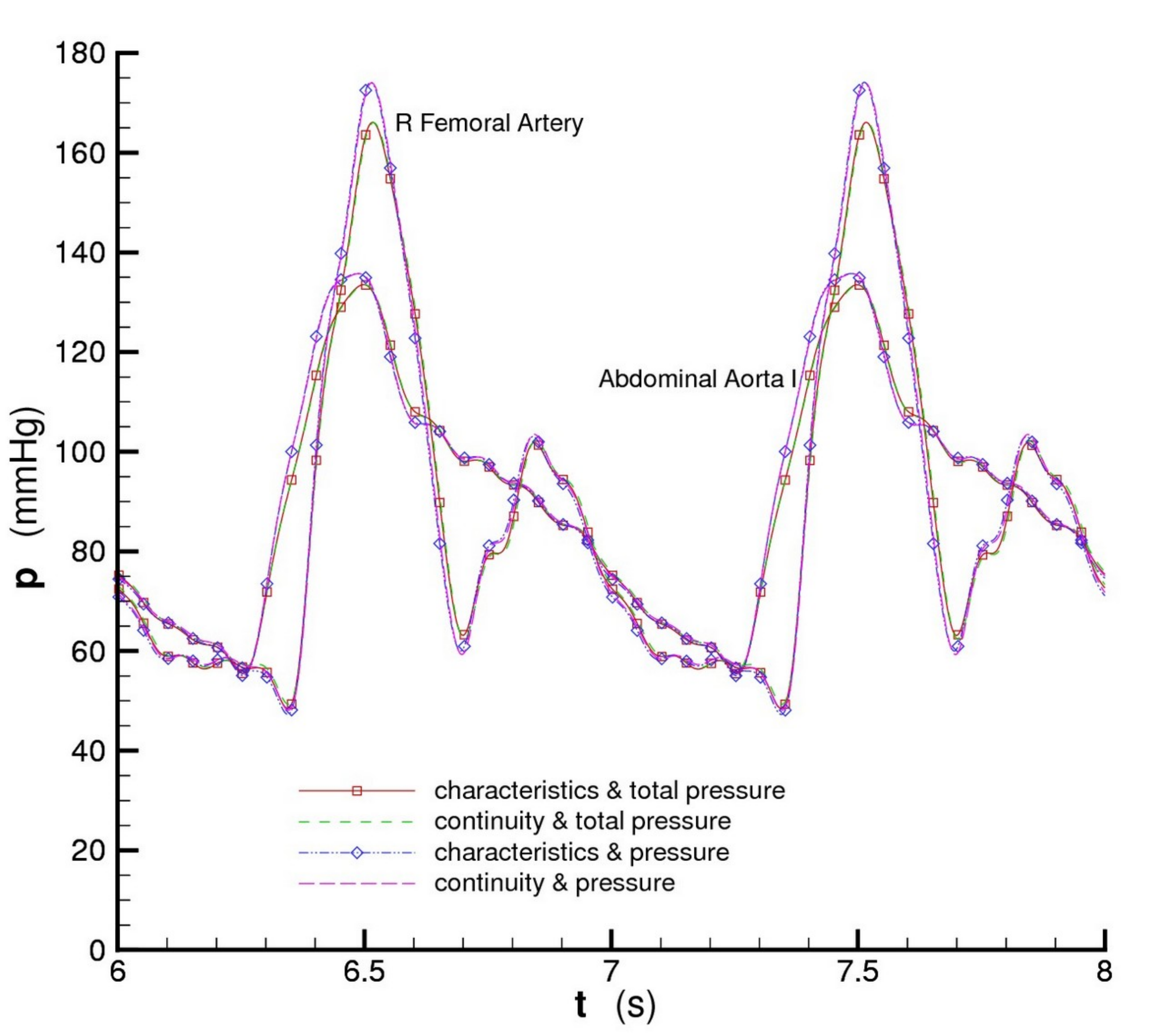}
\caption{Comparison of the bifurcation point treatments summarized in Table 2 by showing the effects on pressure waveforms.}
\label{fig:obifp}
\end{figure}

\begin{figure}
\centering
\includegraphics[width=0.7\textwidth]{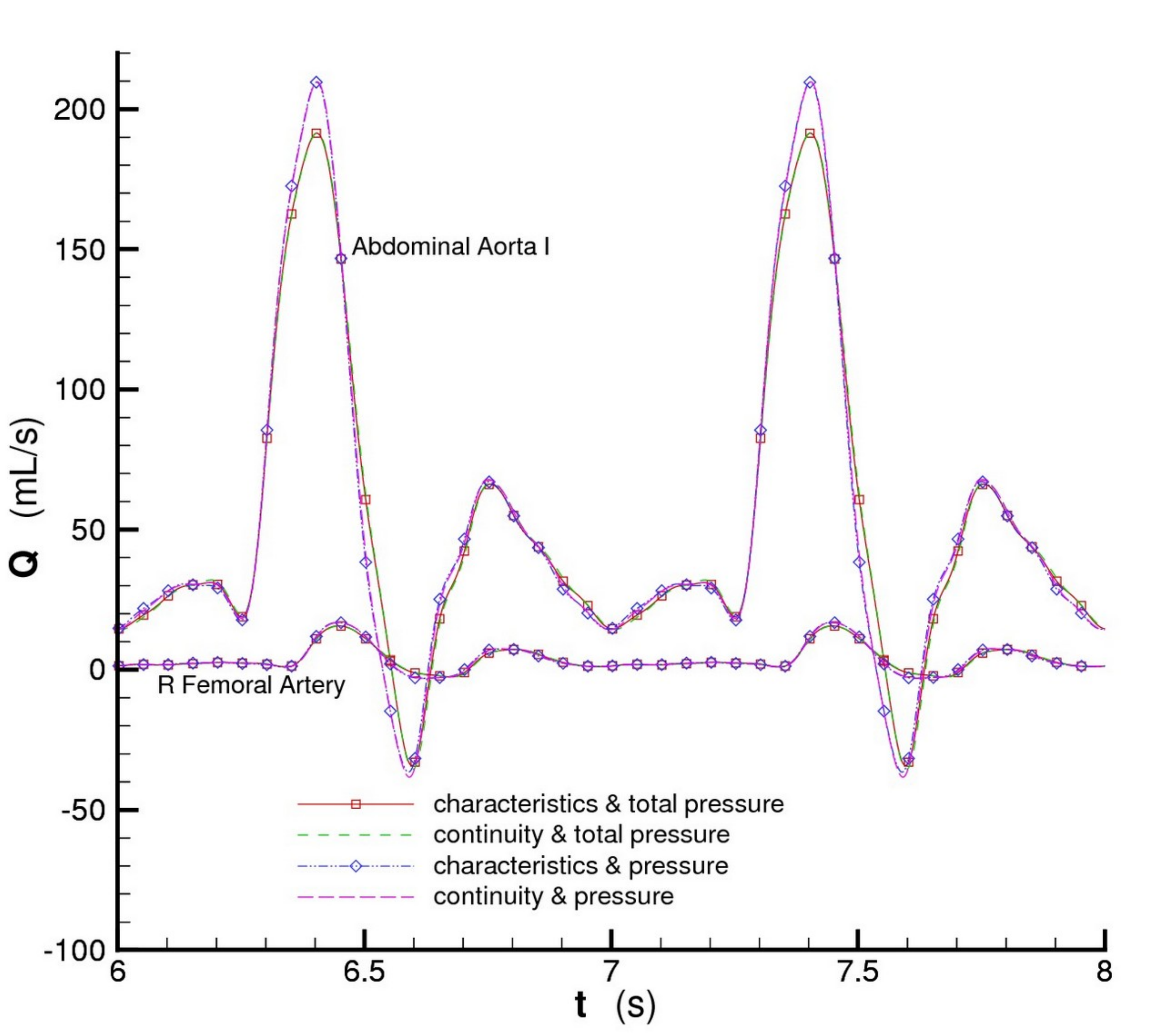}
\caption{Comparison of the bifurcation point treatments summarized in Table 2 by showing the effects on flow waveforms.}
\label{fig:obifq}
\end{figure}

In previous studies, the boundary conditions at the bifurcating points in the network are mostly concentrated on the conservation of mass flow rate at that point along with the conservation of pressure or total pressure. Additional equations to close the nonlinear system of equations at those points come from either Riemann invariants or discrete continuity equations as explained before. Therefore we compared these four strategies to quantify the relative modeling difference. The computed pressure waveforms for Abdominal Aorta I and Right Femoral Arteries are shown in Fig.~\ref{fig:obifp}. Similarly, computed flow waveforms are also illustrated in Fig.~\ref{fig:obifq}. The results demonstrate that the modeling with pressure or total pressure strategies at the bifurcation points effects the waveforms considerably. However, the complementary equations coming from the Riemann invariants or discrete continuity equations does not effect the results.
\subsection{Effects of the outflow boundary conditions}
\begin{figure}
\centering
\includegraphics[width=0.7\textwidth]{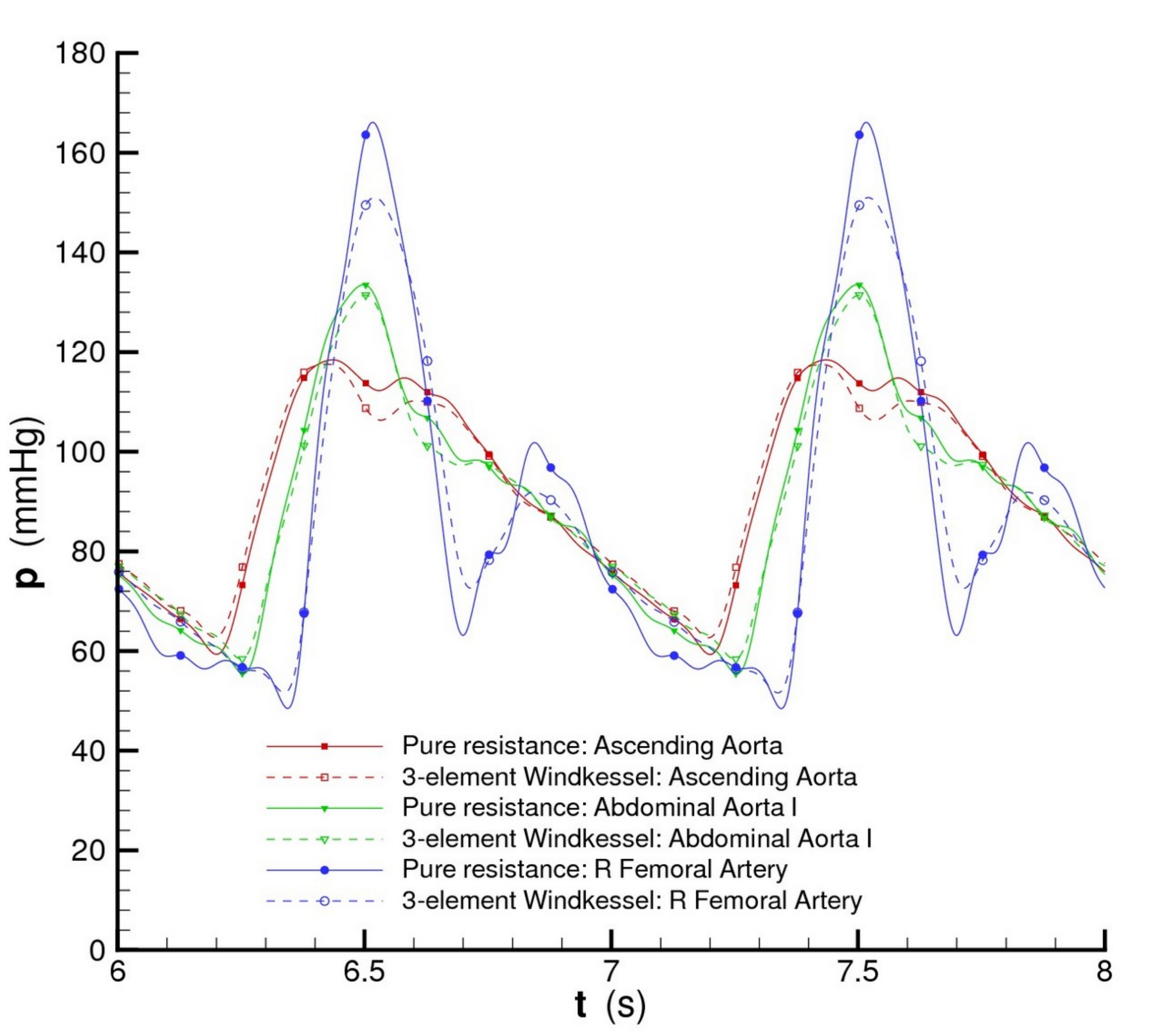}
\caption{Comparison of the lumped parameter terminal boundary conditions by showing the effects on pressure waveforms.}
\label{fig:outp}
\end{figure}

\begin{figure}
\centering
\includegraphics[width=0.7\textwidth]{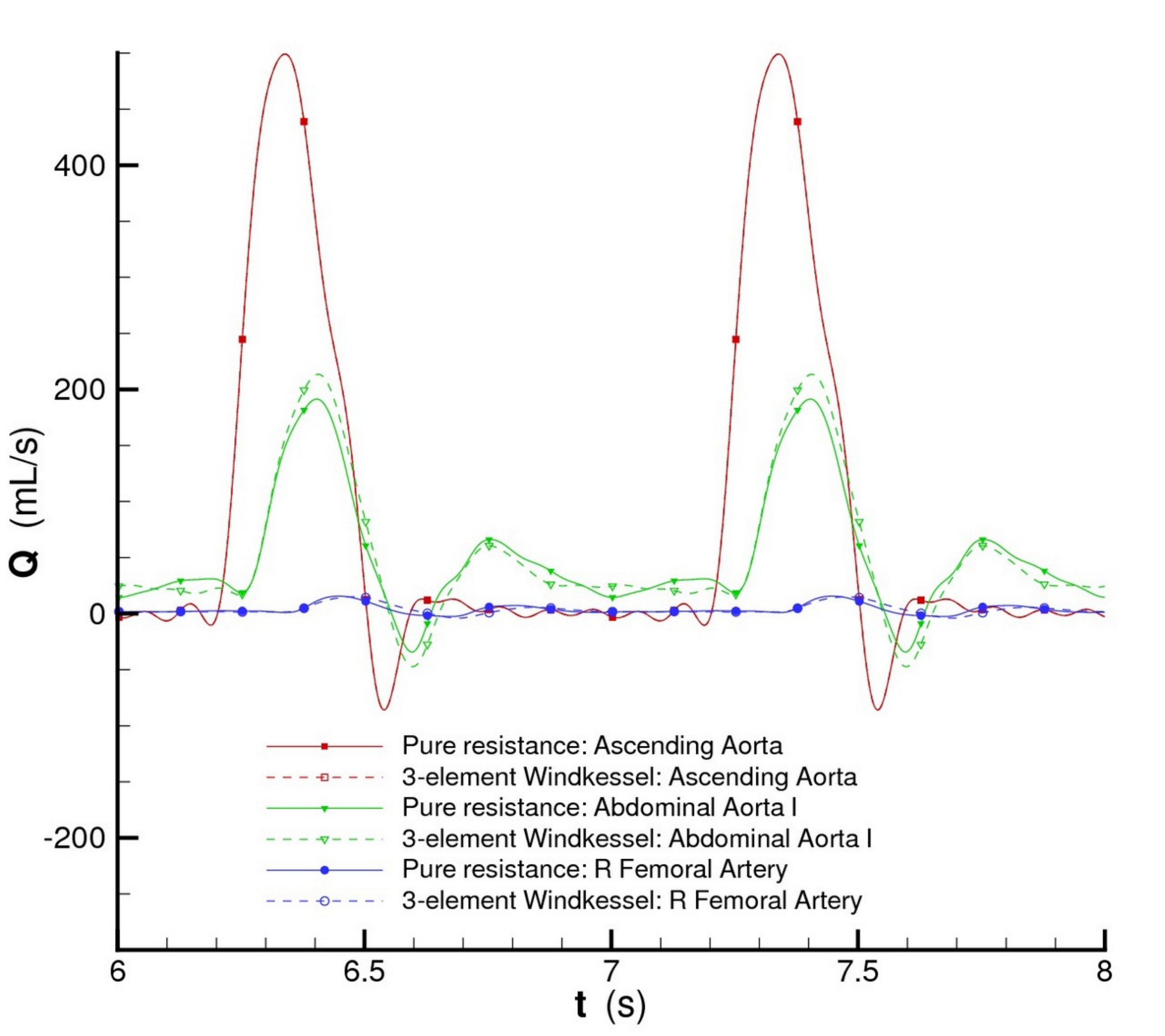}
\caption{Comparison of the lumped parameter terminal boundary conditions by showing the effects on flow waveforms.}
\label{fig:outq}
\end{figure}

The imposed distal boundary conditions depend on the type of lumped parameter model used at the terminal points.
Various lumped parameter outflow boundary condition models including Eq.~\ref{eq:3WK} and Eq.~\ref{eq:wkres} are available for arterial tree simulations \cite{grinberg2008outflow,olufsen2004deriving,vignon2004outflow}. When we analyzing the effects of certain modeling parameters in previous sections, we have used the pure resistance model. Here, we compared the three-element Windkessel and pure resistance models for the terminal points in order to see the compliance effects beyond the terminal points. The physiological data for terminal segments, including the $R_T$ and $C_T$ values, was tabulated in \cite{stergiopulos1992computer,wang2004wave,alastruey2009analysing}, and the constant value of $R_1/R_T = 0.2$ was used for Windkessel outflow boundary conditions. Therefore, the biggest part of total resistance (80\%) is attributed to the effect of the small vessel at the distal microvasculature. The compliances of distal arteries ($16.5\%$) were estimated to give a total arterial compliance of about 1.0 mL/mmHg. In order to show the effects of these two models, the pressure and flow waveforms are compared at three locations. As shown in Fig.~\ref{fig:outp} and Fig.~\ref{fig:outq}, the treatment of outflow boundary conditions have significant effects for modeling of arterial tree.

\section{Conclusion}
\label{sec:conc}
An improved model has been developed for the one-dimensional theory of flows through elastic tubes based on transient Womersley profiles changing continuously via the shape parameter. The model equations, the Pulsed Flow Equations (PFE), are a set of coupled partial differential equations that capture features particularly relevant to internal flows through flexible elastic tubes, such as flows in respiratory and circulatory systems. The equations are an extension of the standard one-dimensional fluid flow equations that are able to capture some additional two-dimensional transport, viscous, and branching effects. The improvement of the one-dimensional theory by introducing functional coefficients to the momentum equation suggests that the formalism provided here can be used as a more appropriate global macroscopic reduced-order distributed mathematical model for the pulsatile fluid dynamics in physiological systems. The PFE are applied to a model of the human arterial tree containing the largest fifty-five arteries to investigate the blood flow characteristics for various velocity profiles and bifurcation modeling strategies, as well as the different lumped parameter outflow boundary conditions. The equations are discretized by the Lax-Wendroff scheme which is second-order accurate in time and space. The results demonstrate that the effect of the assumed velocity profile are significant for arterial tree modeling. The results also show that both the bifurcation point modeling and the treatment of the terminal boundary conditions have also significant effects on the results. The quadratic nonlinear velocity term in the momentum equation also has a minor importance in modeling. Finally, the formalism is not restricted to the human cardiovascular system. It can also be used to compute other internal flows in flexible channels, including other physiological flows. Flows in the respiratory system, for example, and can be used as an efficient tool for computing the reduced-order macroscopic level of description of geometric multiscale models.

\section*{Acknowledgments}
AES extends her gratitude to J.P. Boris, E.S. Oran, C. Kaplan and K. Kailasanath of the Laboratory for Computational Physics and Fluid Dynamics (LCPFD) at the Naval Research Laboratory in Washington DC for fostering this work, and for their many invaluable suggestions. She also acknowledges the Defense Threat Reduction Agency, which partially funded this work under Project Number B082617M, and the Postdoctoral Research Associate program of the National Academies National Research Council, which also partially funded this work while AES was in residence at LCPFD.

\section*{Appendix}
\label{sec:appendix}
Since the contribution of the viscous dissipation term, $\nu\frac{\partial^2 \bar{u}}{\partial x^2}$, to the momentum equation is negligible, the governing PFE's given in Eq.~\ref{eq:GE} can be considered as a coupled hyperbolic system. The characteristics analysis of this system can also be performed by writing the system in non-conservative form. When $A_0$ and $c_0$ are constant and $\alpha=1$, we can approximate this hyperbolic system in non-conservative form as:
\begin{equation}
\frac{\partial Q}{\partial t} + H \frac{\partial Q}{\partial x} = S
\label{eq:nonc}
\end{equation}
where
\begin{equation}
H=\frac{\partial F}{\partial Q} = \left[ \begin{array}{cc}
\bar{u} & A  \\
c_{0}^{2}\sqrt{A_{0}} A^{-3/2} & \bar{u} \end{array} \right]
\end{equation}
The eigenvalues of $H$ are the roots of the characteristic equation $|H-\lambda I|$. There are two distinct real eigenvalues $\lambda^{+} = \bar{u} + c_{0}(\frac{A_{0}}{A})^{1/4}$, and $\lambda^{-} = \bar{u} - c_{0}(\frac{A_{0}}{A})^{1/4}$, and hence the system is totally hyperbolic. The left eigenvectors are non-trivial solutions of the system:
\begin{equation}
\left[ \begin{array}{cc} l_1 & l_2  \end{array} \right]  \left[ \begin{array}{cc} \bar{u}-\lambda & A  \\
c_{0}^{2}\sqrt{A_{0}} A^{-3/2} & \bar{u}-\lambda \end{array} \right] =  \left[ \begin{array}{cc} 0 & 0  \end{array} \right]
\label{eq:1}
\end{equation}
Hence, the solutions are
\begin{equation}
l_{1}^{\pm}=c_{0} A_{0}^{1/4}A^{-5/4}l_{2}^{\pm}
\label{eq:1}
\end{equation}
where $l_{2}^{\pm}$ is arbitrary. To obtain the Riemann invariants, $q$, we consider the following definitions:
\begin{equation}
\frac{\partial q^{\pm} }{ \partial A} = \mu^{\pm} l_{1}^{\pm}, \quad \frac{\partial q^{\pm} }{ \partial u} = \mu^{\pm} l_{2}^{\pm}
\label{eq:r1}
\end{equation}
The integrating factor, $\mu$, is determined from the consistency condition:
\begin{equation}
\frac{\partial (\mu^{\pm} l_{1}^{\pm} )}{ \partial u} =\frac{\partial (\mu^{\pm} l_{2}^{\pm}) }{ \partial A}
\end{equation}
By choosing $l_{2}^{\pm}=1$, and $\mu^{\pm}=1$, Riemann invariants which obtained by integrating the two equations given in Eq.~(\ref{eq:r1}) implying the initial conditions of $\bar{u}=0$ at $A=A_0$, are given as:
\begin{equation}
q^{\pm} = \bar{u} \pm 4 c_0 [1-(A_0/A)^{1/4}]
\end{equation}
Finally, the Riemann invariant form of the non-conservative hyperbolic system is given as
\begin{equation}
\frac{\partial  }{ \partial t} (\bar{u} \pm 4 c_0 [1-(A_0/A)^{1/4}]) + \lambda_{\pm} \frac{\partial  }{ \partial x} (\bar{u} \pm 4 c_0 [1-(A_0/A)^{1/4}]) = -\frac{1}{\rho}\frac{\partial p_0}{\partial x} -\beta \pi \nu\frac{\bar{u}}{A}
\label{eq:rie}
\end{equation}
When the right-hand-side of the Eq.~(\ref{eq:rie}) goes zero, $dq^{-} = 0$ on the characteristic line $C^{-}: \frac{dx}{dt}=\lambda^{-}$, and $dq^{+} = 0$ on the characteristic line $C^{+}: \frac{dx}{dt}=\lambda^{+}$. Therefore, we can use these approximate Riemann invariants when we construct the boundary conditions.


\bibliographystyle{unsrtnat}
\bibliography{ref}

\end{document}